\documentclass[twocolumn,tighten,longauthor]{aastex61}

\usepackage{natbib}
\usepackage{color}
\usepackage{pslatex}
\usepackage{url}
\usepackage{lineno}

\newcommand{\lya}{Ly$\alpha$}
\newcommand{\lyc}{LyC}

\newcommand{\hst}{{\em HST}\/}

\newcommand{\jwst}{{\em JWST}\/}

\newcommand{\galex}{{\em GALEX}\/}

\newcommand{\flux}{erg cm$^{-2}$ s$^{-1}$ \AA$^{-1}$ }
\def\cmtwosa{\ifmmode {\rm cm^2\;s\;\AA} \fi}
\received{26 May 2017}
\revised{29 June 2017}
\accepted{11 July 2017}
\submitjournal{ApJ}
\shorttitle{Requirements for the detection of Lyman continuum}
\shortauthors{McCandliss, O'Meara}

\begin{document}

%% LaTeX will automatically break titles if they run longer than
%% one line. However, you may use \\ to force a line break if
%% you desire.

\title{Flux Sensitivity Requirements for the Detection of Lyman Continuum Radiation Drop-ins from Star-forming Galaxies Gelow Redshifts of 3}

%% Use \author, \affil, and the \and command to format
%% author and affiliation information.
%% Note that \email has replaced the old \authoremail command
%% from AASTeX v4.0. You can use \email to mark an email address
%% anywhere in the paper, not just in the front matter.
%% As in the title, use \\ to force line breaks.

\correspondingauthor{Stephan R. McCandliss}
\email{stephan@pha.jhu.edu}

\author{Stephan R. McCandliss} 
\affiliation{Center for Astrophysical Sciences, Department of Physics and Astronomy, The Johns Hopkins University, Baltimore, MD  21218, USA}

\author{John M. O'Meara}
\affiliation{Department of Physics, Saint Michael's College, Colchester, VT  05439, USA}

%% Notice that each of these authors has alternate affiliations, which
%% are identified by the \altaffilmark after each name.  Specify alternate
%% affiliation information with \altaffiltext, with one command per each
%% affiliation.

%% Mark off your abstract in the ``abstract'' environment. In the manuscript
%% style, abstract will output a Received/Accepted line after the
%% title and affiliation information. No date will appear since the author
%% does not have this information. The dates will be filled in by the
%% editorial office after submission.

\begin{abstract}
Flux estimates for ionizing radiation escaping from star-forming galaxies with characteristic UV luminosities ($L^{*}_{1500(1+z)}$) derived from \galex\ and the VIMOS-VLT Deep Survey, are presented as a function of redshift and assumed escape fraction.  These estimates offer guidance to the design of instrumentation and observing strategies, be they spectroscopic or photometric, for attempting to detect \lyc\ escaping star-forming galaxies for redshifts $ z < 3$.  Examples are given that relate the integrated escape fraction ($f^{e}_{LyC}$) of ionizing photons, obtained by integrating over the entire extreme UV (EUV) bandpass, to the relative escape fraction ($f^{e}_{900}$) observed just shortward of the ionization edge at 911.8 \AA\ as a function of \ion{H}{1}, \ion{He}{1}, and \ion{He}{2}  column densities.  \added {We find that for $\log{N_{HI}(cm^{-2})} \ga 17.0$ $f^{e}_{LyC}$ is significantly greater than $f^{e}_{900}$.}  Detection of \lyc\ ``drop-ins'' in the rest-frame EUV will provide enhanced fidelity to determinations of the integrated fraction of ionizing photons $f^{e}_{LyC}$ that escape  star-forming galaxies and contribute to the metagalactic ionizing background (MIB).  
\end{abstract}

%% Keywords should appear after the \end{abstract} command. The uncommented
%% example has been keyed in ApJ style. See the instructions to authors
%% for the journal to which you are submitting your paper to determine
%% what keyword punctuation is appropriate.


%% From the front matter, we move on to the body of the paper.
%% In the first two sections, notice the use of the natbib \citep
%% and \citet commands to identify citations.  The citations are
%% tied to the reference list via symbolic KEYs. The KEY corresponds
%% to the KEY in the \bibitem in the reference list below. We have
%% chosen the first three characters of the first author's name plus
%% the last two numeral of the year of publication as our KEY for
%% each reference.

%% Authors who wish to have the most important objects in their paper
%% linked in the electronic edition to a data center may do so by tagging
%% their objects with \objectname{} or \object{}.  Each macro takes the
%% object name as its required argument. The optional, square-bracket 
%% argument should be used in cases where the data center identification
%% differs from what is to be printed in the paper.  The text appearing 
%% in curly braces is what will appear in print in the published paper. 
%% If the object name is recognized by the data centers, it will be linked
%% in the electronic edition to the object data available at the data centers  
%%
%% Note that for sources with brackets in their names, e.g. [WEG2004] 14h-090,
%% the brackets must be escaped with backslashes when used in the first
%% square-bracket argument, for instance, \object[\[WEG2004\] 14h-090]{90}).
%%  Otherwise, LaTeX will issue an error. 

\section{Introduction}
\label{intro}

It is evident that most of the hydrogen in the universe was reionized during an Epoch of Reionization (EOR) somewhere between 13.4 and 12.7 Gyr ago, corresponding to redshifts  12 $> z >$ 6 \citep{Fan:2006, Bouwens:2015}, when primordial gas clouds began to collapse into proto-galaxies to form the first stars and black holes.  Whether the overall increase in ionizing radiation that precipitated the EOR was produced by the first stars or black holes is a major unanswered cosmological question.  
 
The total budget for ionizing radiation escaping from these objects remains uncertain \citep{Madau:2015}, but its history plays a crucial role in regulating the subsequent emergence and evolution of structure in the universe \citep[c.f.][]{Madau:1999, Ricotti:2002, Benson:2013, Robertson:2015}.  Lyman continuum (LyC) photons, emitted below the rest frame \ion{H}{1} ionization edge at 911.8 \AA, escape the highly ionized confines of quasars and active galactic nuclei (AGNs) with relative ease \citep{Bahcall:1967, Smith:1981, Bechtold:1987, Scott:2004}, but the potential contribution from the vastly more numerous star-forming galaxies is harder to quantify; this is due to the difficulty of observing this intrinsically weak emission coupled with our poor understanding of the physical conditions that allow ionizing radiation to escape into the intergalactic medium (IGM).

Empirical estimates of the average escape fraction required to sustain an ionized IGM by $z$ = 6, range from  5 $ < f^e_{LyC} < $  40\% \citep[][and references therein]{Bouwens:2015, Finkelstein:2015}.  These estimates depend on the evolution of the steepness of the galaxy luminosity function, the mean production of LyC emission per unit star-formation rate (SFR), and assumptions regarding the ratio of the escape fraction to the ionized hydrogen clumping factor \citep{Madau:1999}.   The conclusion that the EOR is driven solely by star-forming galaxies rests on these assumptions, and on an extrapolation of the faint-end cutoff of the galaxy luminosity function from -17 $ < M_{uv} < $ -13. Moreover, there are no constraints as to how the escape fraction is distributed as a function of galactic mass, luminosity, and environment.  

Cosmological hydrodynamical simulations aiming to determine the escape fraction as a function of luminosity and halo mass have produced mixed results \citep[c.f.][]{Gnedin:2008, Razoumov:2010, Yajima:2011, Wise:2014, Yajima:2014}.  In recent work, \citet{Sharma:2016} found that the brighter galaxies have higher escape fractions.  In contrast,  \citet{Xu:2016} find that galaxies with smaller halo masses have the highest escape fractions, however, they also found that star formation in low-mass galaxies is easily suppressed as reionization progresses, leaving higher-mass galaxies, which are less susceptible to photo-evaporation, to complete the process.  

One of the key projects for the James Webb Space Telescope (\jwst) is to search for those sources responsible for reionizing the universe; however, it will likely only be able to do so indirectly.  The monotonic increase with redshift in the density of Lyman limit systems (LLS) -- those discrete clouds in the IGM having  $\log(N_{HI}(cm^{-2})) > $ 17.2 --  steadily decreases the probability of directly detecting LyC emission  from star-forming galaxies on an unattenuated line of sight \citep[c.f.][]{Madau:1995, Inoue:2008, Inoue:2014, Worseck:2014, Crighton:2015}.  At redshifts of $z$ = [3, 4, 5, 6] the mean transmission of the IGM is estimated to be $\approx$ [0.5, 0.3 0.08, 0.01] \citep[][their Figure 4]{Inoue:2014}, albeit with an large variation about the mean.   The \jwst\  short wavelength cutoff is $\approx$ 0.6 $\mu$m, so the LyC region is accessible only for redshifts $z \ga$ 6 where the IGM is essentially completely opaque.  Direct measurements of LyC will be a challenge for \jwst, to say the least

The far-UV and near-UV bandpasses provide the most direct path to spatially resolved detection of ionizing radiation, and to the characterization of those environments that favor LyC escape.  Spatial resolution is an especially important diagnostic, as models indicate that LyC photons escaping from any particular galaxy will exhibit gross variations that depend on the line of sight of a star-forming source with respect to intervening neutral and ionized material in disks, superbubbles, and surrounding circumgalactic streams  \cite{Dove:1994, Bland-Hawthorn:1999, Dove:2000, Shull:2015}. 

Recent successes in detecting LyC emission from what are apparently star-forming galaxies \citep{Leitet:2013, Borthakur:2014, Izotov:2016b, Leitherer:2016, Naidu:2016, Shapley:2016} have emboldened the design of instrumentation and observing strategies capable of quantifying, on an industrial scale, the relative contributions of star-forming galaxies, quasars, and AGN to the creation and sustenance of the metagalactic ionizing background (MIB) across cosmic time.

The need for understanding those physical processes that enable $f^e_{LyC}$ at low redshift has grown in importance as of late.  Recent determinations of the number density of \lya\ forest lines found at low redshift by  \citet{Danforth:2016} appear to require a MIB $\sim$ 5$\times$ larger than theoretical estimates to explain the low density of the lines \citep{Kollmeier:2014};   \citep[see][for contrasting conclusions]{Shull:2015, Gaikwad:2017}.  These studies are inconclusive as to whether, on average, $f^e_{LyC}$ from star-forming galaxies and quasars is considerably higher than the handful of detections to date indicate, thus underlining the importance of quantifying $f^e_{LyC}$ at both low and high redshift.  

Our main goal is to establish effective area requirements for future observatories that will transform what has previously been described as an impossible task \citep{Fernandez-Soto:2003}, into a statistically significant determination of LyC luminosity function evolution across cosmic time envisioned by \citet{Deharveng:1997} and \citet{Shull:2015}, providing a full accounting of the LyC escape budget from star-forming galaxies of all types.  

Here we use 1500 \AA\ luminosity functions  to guide estimates of the rest frame EUV flux (the LyC) emitted by characteristic galaxies,  attenuated by a uniform foreground screen of circumgalactic media (CGM) with representative ratios of \ion{H}{1}, \ion{He}{1}, and \ion{He}{2} column densities.   We also include the progressive increase in mean attenuation effected by the increasing \ion{H}{1} distribution of IGM column densities as a function of redshift.  A general result is that the LyC escape fraction measured in a narrow range just shortward of the ionization edge is a poor representation of the total fraction of ionizing photons that escape.  

These calculations were developed to support science and technology flowdown exercises for the Large Ultra-Violet Optical InfraRed (LUVOIR) and Habitable Exoplanet (HabEx) survey mission studies commissioned by NASA in preparation for the Astrophysics Decadal Survey for 2020, and to assess the capability of proposed probe and explorer class missions. 

A cosmology of $H_0 = $ 70 km s$^{-1}$ Mpc$^{-1}$, $\Omega_m =$ 0.3,   $\Omega_{\Lambda} =$ 0.7 is assumed.

\subsection{Review of Escape Fraction Terminology}

The term  ``escape fraction'' is often used loosely and somewhat conflictingly.  The literature defines a number of slightly different ratios to describe the escape of ionizing radiation from galactic environments, which for the sake of completeness we review briefly.

A useful observable, termed the relative escape fraction ($f_{rel}$) by \citet{Reddy:2016, Shapley:2006}, and \citet{Steidel:2001},  is defined as the ratio of the observed flux in a bandpass just shortward of the Lyman edge ($F^o_{900}$) to that observed at some fiducial UV continuum wavelength ($F^o_{\lambda}$);
\begin{equation}
f_{rel} = F^o_{900}/F^o_{\lambda}.
\end{equation}

\noindent The observed flux is then relatable to the intrinsic flux ($F^i_{\lambda}$) by appeal to a stellar population spectral synthesis model, after accounting for the transmission of flux, $T_{\lambda}$, as attenuated by line of sight neutral gas and dust, such that $F^o_{\lambda} =T_{\lambda} F^i_{\lambda}$.  Just over the edge we have $F^o_{900} =f^e_{900} F^i_{900}$, which  
\citet{Reddy:2016} called $f^e_{900}$  the absolute escape fraction (in their notation $f(LyC)_{abs}$).

Solving for $f^e_{900}$ in terms of the relative escape fraction, the attenuation at the fiducial wavelength and the intrinsic flux ratio,
 \begin{equation}
f^e_{900} = T_{\lambda} f_{rel} F^i_{\lambda}/F^i_{900}.
\end{equation}

\noindent  It explicitly includes hydrogen and dust attenuation at the edge.   \citet{Borthakur:2014} made a further distinction between what they called a relative escape fraction at 912$^{-}$, where $T_{\lambda}$ =1, and an absolute escape fraction at 912$^{-}$ where they corrected for dust using the ratio of the integrated LyC luminosity, with respect to the bolometric luminosity.

Implicit in these definitions is the assumption that the observed flux $F^o_{\lambda}$ is averaged over some relatively narrow bandpass.  However, $f^e_{900}$ does not fully account for all the ionizing photons that escape from the CGM and contribute to the MIB.  To do so requires an integration of the attenuated (``observable'') photon number flux ($F^o_{\lambda}/E_{\lambda}$; where $\lambda <$ 911.8 \AA\ and $E_{\lambda}$ is the photon energy)  over the full extent of the spectral energy distribution (SED) shortward of the Lyman edge.  The integrated escape fraction is normalized by the intrinsic SED integrated over the same wavelength interval,

\begin{equation}
f^e_{LyC}= \frac{\int_{0}^{912}F^o_{\lambda}/E_{\lambda} d\lambda}{\int_{0}^{912}F^i_{\lambda}/E_{\lambda} d\lambda}.
\end{equation}

\noindent In practice, $F^o_{\lambda}$ is not observable in its entirety; however, we can estimate it by the integration of a transmission function, multiplied by the intrinsic flux, $T_{\lambda}F^i_{\lambda}$, over the wavelength interval shortward of the Lyman edge.

In \S~\ref{lumfun} we provide estimates of $F^o_{(1+z)900}$ in terms of the characteristic apparent ab-magnitude (abmag) $m_{(1+z)900}$,  as a function of  $f^e_{900}$, scaled from compilations of UV luminosity functions taken from the literature for rest frame wavelengths $\approx$ 1500 \AA.   In \S~\ref{dropins} we will give estimates of the integrated fraction of escaping ionizing photons, $f^e_{LyC}$, by modeling the optical depth in neutral hydrogen, neutral helium, and once-ionized helium column in the CGM in the foreground of a spectral synthesis model.  In \S~\ref{igmatten} we include the IGM attenuation in estimating the LyC flux of redshifted models scaled to the magnitudes of characteristic galaxies at (1+$z$)1500 \AA, and  discuss detection requirements and observing strategies.

\begin{deluxetable}{cccccc}
\tablecaption{\bf Schechter parameters for UV Luminosity Functions \label{t1} }
\tablecolumns{6}
\tablewidth{0pt}
\tablehead{\colhead{$z$}	&  \colhead{$\Delta z$}	&\colhead{$\alpha$} &	\colhead{$M^*_{15}$} & \colhead{$\phi^{*}$\tablenotemark{a} }  &  \colhead{$m^*_{(1+z)1500}$} }
\startdata
0.1& 0 - 0.2		& -1.21	& -18.05 	& 4.07	& 20.17\\
0.3& 0.2 - 0.4	& -1.19	& -18.38	& 6.15	& 22.30\\
0.5& 0.4 - 0.6	& -1.55	& -19.49	& 1.69	& 22.37\\
0.7& 0.5 - 0.8	& -1.60	& -19.84	& 1.67	& 22.73\\
1.0& 0.8 - 1.2	& -1.63	& -20.11	& 1.14	& 23.24\\
2.0& 1.75 - 2.25	& -1.49	& -20.33	& 2.65	& 24.43\\
2.9& 2.4 - 3.4	& -1.47	& -21.08	& 1.62	& 24.38\\
\enddata
\tablenotetext{a}{(10$^{-3}$ Mpc$^{-3}$ \added{mag$^{-1}$} )}
\end{deluxetable}

\section{Far-UV Luminosity Functions}\label{lumfun}

We  use far-UV luminosity functions listed by \citet{Arnouts:2005} to estimate areal density of candidates and their corresponding LyC flux as functions of assumed escape fraction and redshift.  In Table~\ref{t1} we list the \citet{Schechter:1976}  function parameters.  \added{The luminosity function in units of absolute magnitude is expressed as}

\begin{equation}
\label{eq0}
\phi(M_{15})=\frac{\ln{10}}{2.5}\phi^*10^{-\frac{(M_{15}-M^{*}_{15})(\alpha+1)}{2.5}}e^{-\frac{(M_{15}-M^*_{15})}{2.5}},
\end{equation}
\explain{The equation has changed.  The $^*$ was dropped on the left side to create the proper independent variable $M_{15}$ and the $M$ in the exponential function on the right was changed to $M_{15}$ to agree with the formula as expressed in Yoshida et al. (2006).}

\begin{figure}
\includegraphics[width=.5\textwidth]{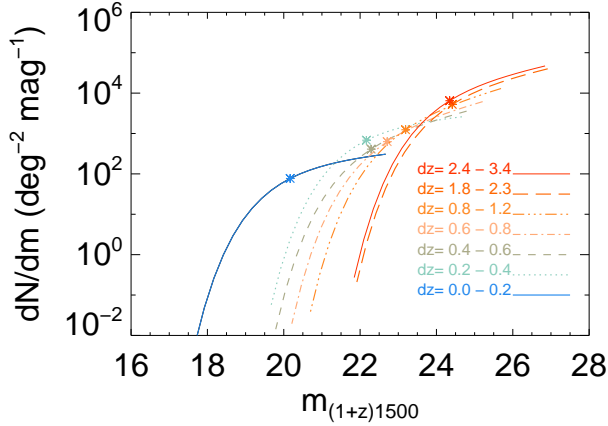}
\caption{Star-forming galaxy surface density as a function of redshift derived from the luminosity functions listed in Table~\ref{t1}.  \label{fig1}}
\end{figure}

\noindent \added{where $M^*_{15}$ is the characteristic absolute magnitude of the Schechter function at 1500 \AA, $\alpha$ is the faint-end power-law slope, $\phi^*$ is the normalization factor (in Mpc$^{-3}$ mag$^{-1}$),  and $M_{15}$ is the independent variable for the absolute magnitude at restframe 1500 \AA.}  

We convert \added{this function} from the number per comoving volume to the number per square degree by multiplying by the comoving volume per solid angle, calculated at \replaced{for}{about} the mid-point in each redshift interval.  \added{The characteristic absolute magnitude in the 1500 \AA\ rest frame was shifted to a characteristic apparent magnitude in the observer's frame using the formula \citep{Yoshida:2006} }
\begin{equation}
\label{eq5}
m^{*}_{(1+z)1500}=M^*_{15}+2.5log{\left[\frac{(d_l(z)*10^5)^2}{(1.+z)}\right]}. 
\end{equation}
 Here $d_l(z)$ is the luminosity distance in Mpc, and $z$ is the redshift.  

In Figure~\ref{fig1} we show the \added{ultraviolet} luminosity functions listed in Table~\ref{t1} recast into  \added{a differential areal density, graphed logarithmically as}  the number of galaxies per unit magnitude per square degree.  Each curve has an $*$ to indicate the location of the characteristic magnitude $m^*_{(1+z)1500}$ where the Schechter function makes the transition from the exponential cutoff in galaxy counts at the bright end of the luminosity function to the power law extension of the faint end.  The interval for the abscissa of each curve spans 5 mag (a factor of 100 in flux) centered on $m^*_{(1+z)1500}$.    The figure is useful for estimating the number of targets within a given patch of sky down to an instrument's brightness limit. 

\added{The estimate for the apparent characteristic magnitude at $(1+z)900$ \AA\ in the observer's frame, $m^*_{(1+z)900}$, is scaled from $m^{*}_{(1+z)1500}$ using}

\begin{equation}
\label{eq2}
m^*_{(1+z)900} = m^{*}_{(1+z)1500}  + \delta m^{1500}_{900} + \delta m_{esc}.
\end{equation}
where \added{the escape fraction just over the ionization edge is} $\delta m_{esc} = 2.5 \log{f^e_{900}}$, \added{and the ratio of the intrinsic rest frame flux at 1500 \AA\ to that at 900 \AA, }
\deleted{We convert the estimates for the 1500 \AA\ rest frame magnitude to LyC magnitudes, using the scale factors} $\delta m^{1500}_{900} = 2.5\log{F^i_{1500}/F^{i}_{900}}$, \added{was} determined from the STARBURST99 \citep[hereafter SB99]{Leitherer:1999, Leitherer:2014} intrinsic SED \replaced{discussed}{described} in \S~\ref{intrinsicspec}.   Conveniently, in the  rotating stellar evolution model \added{that we have adopted}, $ F^i_{1500}/F^i_{900} \approx$ 1.  \replaced{For constant star-formation this ratio is relatively insensitive to age.}{This ratio is relatively insensitive to age in constant star-formation models.}   In the older, non-rotating SB99 models the range is 1.5 $ \lesssim F^{i}_{1500}/F^{i}_{900} \lesssim$ 3 for ages 10 -- 900 Gyr.  A factor of 2 in the flux ratio amounts to $\delta m^{1500}_{900} = 2.5\log{F^i_{1500}/F^{i}_{900}} \approx 0.75$.   

\begin{figure}
\includegraphics[bb= .75in 4.75in 8in 10.5in, clip=false, width=.5\textwidth]{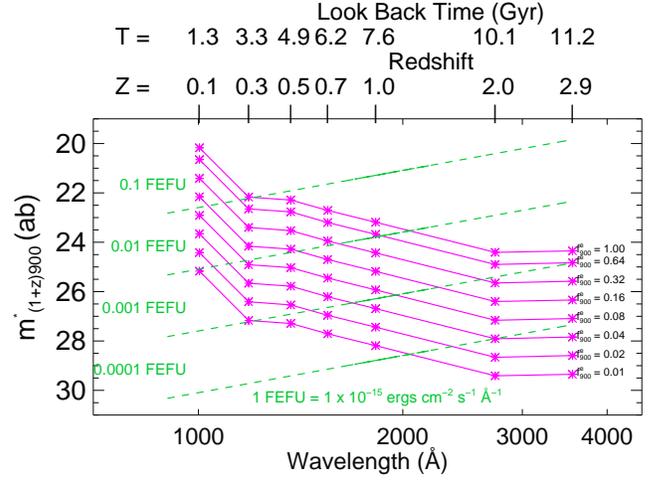}
\caption{Estimate of each luminosity function's characteristic magnitude,  $m^*_{(1+z)900}$, as a function of redshift and assumed  $f^e_{900}$ escape fraction just shortward of the edge Lyman edge.  These estimates do not include IGM attenuation. \label{fig2}}
\end{figure}

The estimates for $m^*_{(1+z)900}$, shown as magenta lines in Figure~\ref{fig2}, are based on the  rotating evolutionary models.  \replaced{The estimates can}{They may} be adjusted to adhere to the older models by shifting the ordinate by the preferred  $\delta m^{1500}_{900}$ (= 0 for the rotational models).   Likewise, offsets to account for differences in dust attenuation between 1500 and 900 \AA\ may also be applied.   

We  have neglected here the stochastic attenuation expected from intervening neutral hydrogen absorption systems associated with the IGM.   We will quantify the effects of this additional attenuation in \S~\ref{igmatten}.

%Updated analytic formulae accounting for mean attenuation as a function of redshift may be found in \citet{Inoue:2014}.  At a redshift of 3 the mean attenuation at $(1+z)900$ is $T_{(1+z)900} \approx$ 0.5 ($\Delta m^*_{(1+z)900} =$ 0.75) and falls precipitously shortward.   

The  estimate for $m^*_{(1+z)900}$ at a given $f^e_{900}$ is intended to provide guidance for detecting LyC emission at the most \replaced{difficult}{attenuated} wavelength, just over the Lyman edge.  The green dashed contours in Figure~\ref{fig2}  indicate the flux levels $F_{(1+z)900}$ in units of femto erg flux units (1 FEFU = 10$^{-15}$ \flux), which was roughly the background limit for the {\it Far Ultraviolet Spectroscopic Explorer (FUSE)}.  Current space-based UV background limits are $\sim$ 1 - 1000 times lower and will be discussed in \S~\ref{future}.

It should be kept in mind that $f^e_{900}$ is not a measure of the ratio of the total number of ionizing photos that escape to the total number that are emitted.  In the following section we show that while the LyC flux in the relatively narrow wavelength range just shortward of the Lyman edge can be essentially zero,  the fraction of escaping continuum  photons integrated over the entire LyC emitting region, from the  edge into the extreme UV (EUV), is significantly greater than zero.

\section{Lyman Drop-ins}\label{dropins}

The abrupt drop-out in flux shortward of the LyC is a useful diagnostic for the photometric identification of star-forming galaxies \citep{Steidel:1995}.   The wavelength of the band where the flux drops out provides a constraint on the redshift.  The technique was originally developed using ground-based surveys, which can efficiently identify objects at redshifts $z \approx$ 3 -- 4.   Lyman drop-outs at these redshifts are commonly referred to as Lyman Break Galaxies (LBGs).   \citet{Cooke:2014} have emphasized that the standard LBG technique is biased toward star-forming galaxies with zero detectable flux in the LyC, i.e. when line of sight column densities of $N_{HI} >> 1 \times $10$^{18}$ cm$^{-2}$.   For $N_{HI} \sim 1 \times $10$^{18}$ cm$^{-2}$ the drop-out does not extend completely throughout the EUV range, so we expect to find a class of star-forming galaxies that we call Lyman ``drop-ins'' as described below.

\subsection{\ion{H}{1}, \ion{He}{1}, \ion{He}{2} CGM transmission model \label{LyCTrans} }

In principal, the shape of the continuum emission shortward of the Lyman edge contains a great deal of information regarding the distribution of ionization states for hydrogen and helium, and the distribution of dust along those unresolved line of sight(s) through an individual galaxy's CGM favoring the escape of LyC radiation.  The attenuation at each photoionization edge increases sharply followed by a relatively gentle recovery, varying approximately as $\propto (\frac{\lambda}{\lambda_e})^3$, where the $\lambda_e$ for \ion{H}{1}, \ion{He}{1}, and \ion{He}{2} are 911.75, 504.26, and 227.84 \AA, respectively.

In general, the transmission below the Lyman edge is a exponential function of the sum of dust, neutral hydrogen, neutral helium, and singly ionized helium optical depths, 
 \begin{equation}
 T_{CGM}(\lambda) = \exp[{-\tau_d(\lambda)-\tau_{HI}(\lambda)-\tau_{HeI}(\lambda)-\tau_{HeII}(\lambda)}],
\end{equation}

\noindent where the various optical depths are \deleted{a} products of the column densities and cross-sections for each species, $\tau_x(\lambda) = N_x\sigma(\lambda)$.  For the hydrogen and helium photoionization cross-sections we use the analytic fits from \citet{Verner:1996}, which differ slightly from the $\lambda^3$ relation.  For the \ion{H}{1}, \ion{He}{1}, and \ion{He}{2} resonance line cross-sections calculations we use the wavelengths and oscillator strengths for the 1 $< n <$ 79 lines of each species found in the online repository complied by Kurucz in CD-ROM 23.\footnote{\url{http://www.cfa.harvard.edu/amp/ampdata/kurucz23/sekur.html}}

We calculated the total optical depth template for each species, following a procedure used by \citet{McCandliss:2003}, wherein Voigt profiles were generated for individual\deleted{s} lines with fine enough sampling and range to resolve the doppler core and return the optical depth in the damping wing to near 0.  Individual line profile optical depths were interpolated onto an common grid spanning the entire wavelength range of interest, for summation with all other lines and photoionization cross-sections.  The doppler velocity was set to $b$ = 35 km s$^{-1}$, as is commonly assumed \citep{Madau:1995, Inoue:2008, Inoue:2014} and is sufficient for our flux estimation exercise; however, it should be kept in mind that observations of the differential distribution of $b$ have been matched to $\frac{dN}{db} = \frac{4b^4_{\sigma}}{b^5}\exp(-(\frac{b_{\sigma}}{b})^4)$ with $b_{\sigma} \approx$ 26 km s$^{-1}$ \citep{Hui:1999}.  How $\frac{dN}{db}$ varies with redshift is an open question.

The relative contributions of the three species are linked through the cosmic abundance of helium by number and imposed assumptions for the neutral fractions with respect to the total for the element (molecular hydrogen is neglected); i.e., $N^{tot}_{H} = N_{HI}+N_{HII}$, $N^{tot}_{He} = N_{HeI}+N_{HeII}+N_{HeIII}$ and $N^{tot}_{He}$ = 0.08 $N^{tot}_{H}$.  The independent variables are the column of \ion{H}{1}, its neutral fraction, $\chi_{HI} \equiv \frac{N_{HI}}{N^{tot}_{H}}$, and the helium neutral fraction, $\chi_{HeI} \equiv \frac{N_{HeI}}{N^{tot}_{He}}$.   

\begin{figure}
\includegraphics[angle=90,width=.5\textwidth]{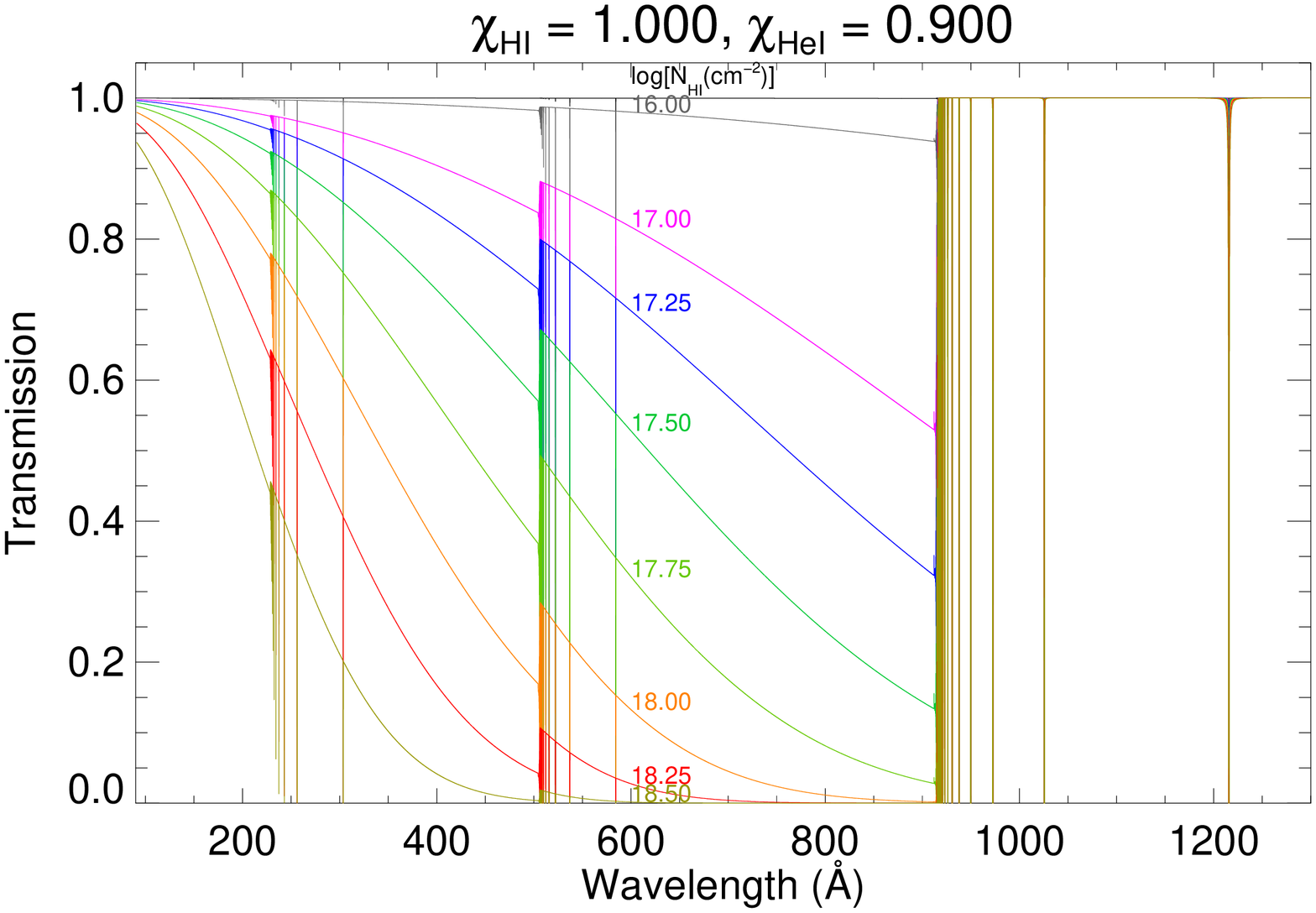}
\includegraphics[angle=90,width=.5\textwidth]{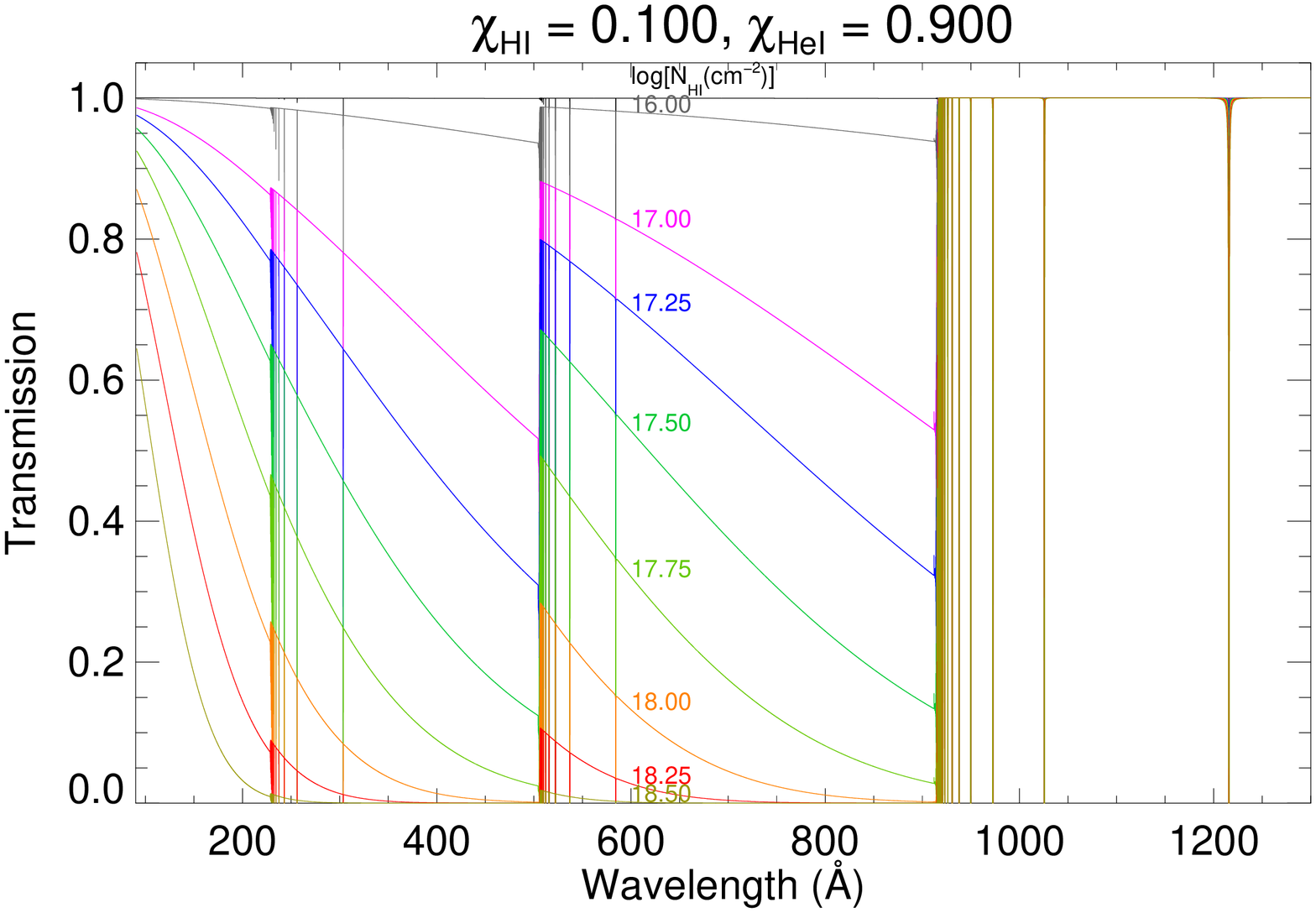}
\caption{LyC transmission functions for $\log{N_{HI}(cm^{-2})}$ = 16.00, 17.00, 17.25, 17.50, 17.75, 18.00, 18.25, 18.50 that span the transition from optically thin to thick at the Lyman edge.  Top --  $\chi_{HI}$ =1, $\chi_{HeI}$ = 0.9.  Bottom --  $\chi_{HI}$ =0.1, $\chi_{HeI}$ = 0.9.\label{fig3}}.
\end{figure}

In Figure~\ref{fig3} we show two sets of transmission models, having ($\chi_{HI}$, $\chi_{HeI}$) = (1.0, 0.9), and (0.1, 0.9), top and bottom, respectively.   The overall blackness of $f^e_{900}$ is set by the $N_{HI}$ column density, but total fraction of escaping LyC photons is set by the ionization state \added{of} hydrogen relative to that of helium.   The strength of the neutral and ionized helium edges becomes enhanced for $\chi_{HI} <$ 1.   

The nature of dust attenuation in a highly ionized density bounded medium below the hydrogen Lyman edge is highly uncertain, so we have neglected it here.  This should not be too serious, as the extinction near the Lyman edge in the canonical Milky Way model of \citet[][their Figure 14]{Weingartner:2001} is $N_{HI}/A(912) \approx$  2.4 $\times$ 10$^{20}$ cm$^{-2}$ mag$^{-1}$.  For $N_{HI}$ = 10$^{18}$ cm$^{-2}$ this yields  $A(912)$ =0.0042, so with $\tau_d(912)= A(912)/1.086$, we estimate a miniscule effect on transmission at the edge of  $\exp{(-\tau_d(912))}$ = 0.996.  However, if the dust abundance follows the total hydrogen instead of just \ion{H}{1}, then $\tau_d$ will be higher.

In principal, a proper accounting of the neutral and ionized gases could lead to important observational insights into the temperature of the CGM along with the survivability and attenuation properties of dust grains with respect to the total gas content in LyC-leaking environments.  

\subsection{Intrinsic SB99 Star Formation Model \label{intrinsicspec} }

We employ here an SB99 spectral synthesis model, forming stars at a continuous rate of 1 $M_{\odot} $yr$^{-1}$ \citep{Leitherer:1999, Leitherer:2014} for the intrinsic SED.  We specified a Kroupa initial mass function, which has double power law exponents of -1.3, over the mass interval from 0.1 $< M_{\odot} <$ 0.5, and -2.3 for 0.5 $< M_{\odot} <$ 100.  We also chose the Geneva evolutionary tracks with rotation at 40\%\ of the break-up velocity (v0.4 models) and solar metallicity were also selected.  After about 10 Myr the number of ionizing photons emitted by this model below the Lyman edge saturates at a rate of $\log{[Q_{0}(s^{-1})]} \sim$  53.4 \citep{Leitherer:1999}.  The mass at this point is 10$^{7} M_{\odot}$.

\subsection{SB99 models with CGM attenuation \label{cgmatten} }

\begin{figure}
\includegraphics[angle=90,width=0.5\textwidth]{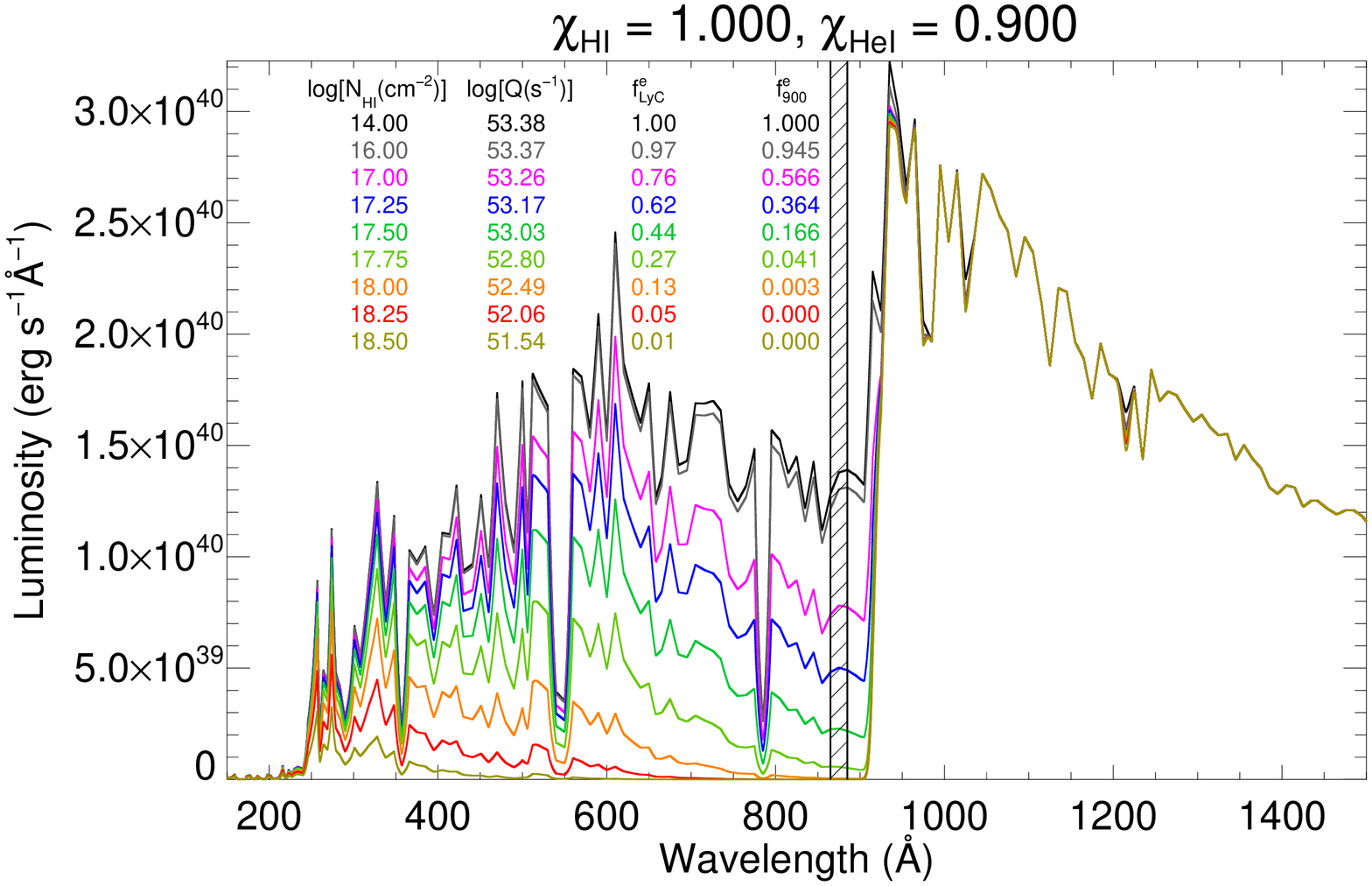}
\includegraphics[angle=90,width=0.5\textwidth]{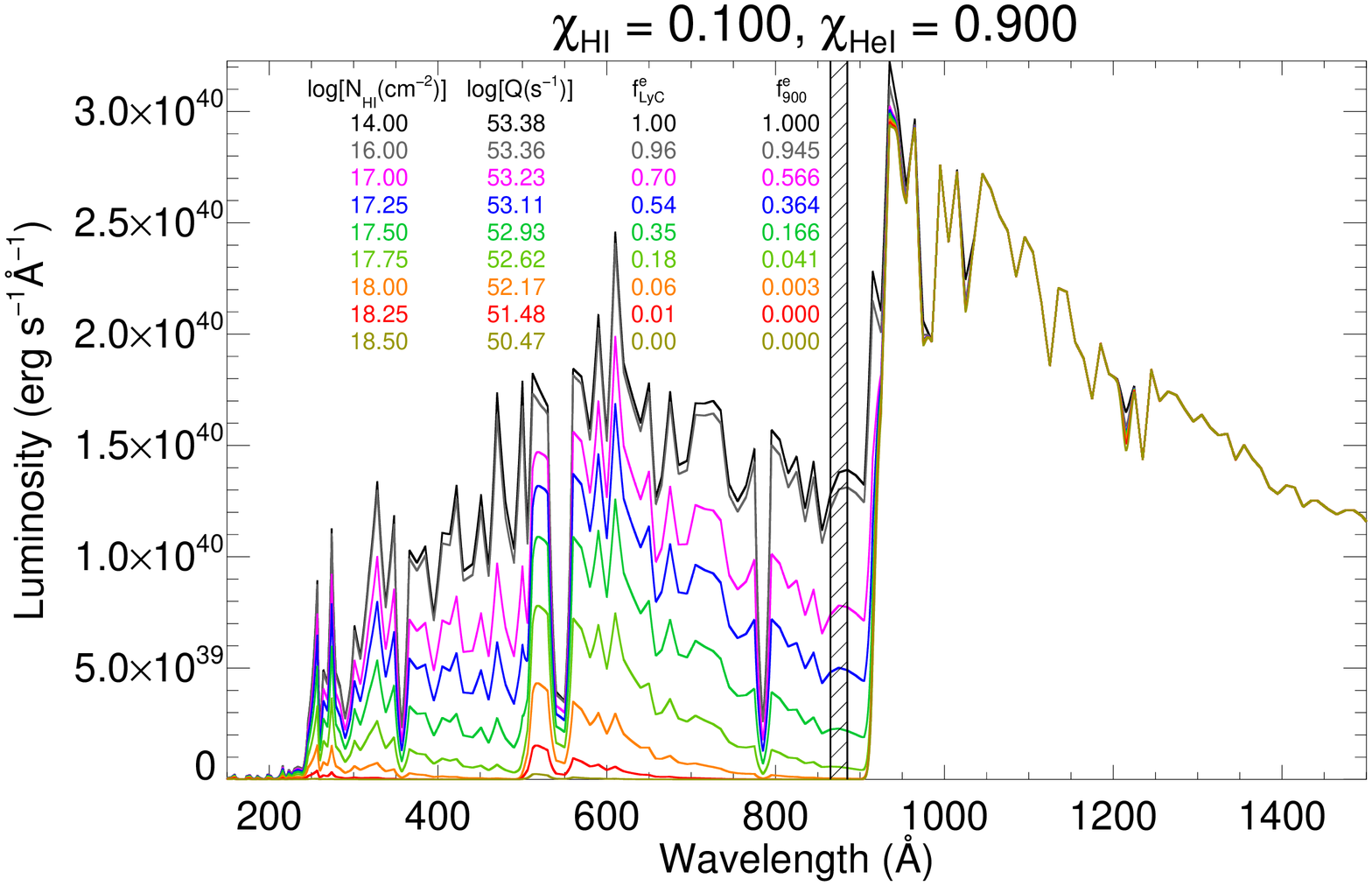}
\caption{LyC drop-ins with ($\chi_{HI}$, $\chi_{HeI}$) as in Figure~\ref{fig3}.  SB99 v0.4 evolution model\deleted{s}  \added{(10 Myr, continuous star formation,  10$^{7} M_{\odot}$),} attenuated with the transmission functions in Figure~\ref{fig3}.   \added{The inset table columns are: $\log{N_{HI}(cm^{-2})}$, the neutral hydrogen column;  $\log{[Q(s^{-1})]}$, the rate of escaping ionizing radiation; $f^e_{LyC}$ = $Q/Q_0$, the integrated escape fraction; and $f^e_{900}$, the escape fraction just shortward of the ionization edge, which is defined as the ratio of the attenuated flux to the intrinsic flux in the $\Delta \lambda$ = 20 \AA\ wide hatched region in the figures.}  For $\log{N_{HI}(cm^{-2})} > $18 the escape fraction measured at 900 \AA\ $f^e_{900} << $1\%, yet the integrated fraction of ionizing photons that escape, $f^e_{LyC}$, is significantly  greater than $f^e_{900}$.  The top panel shows a relatively smooth Lyman drop-in at $\log{N_{HI}(cm^{-2})} = $18.  The bottom panel shows a double drop-in. \label{fig4}}
\end{figure}

\begin{figure}
\includegraphics[angle=90,width=0.47\textwidth]{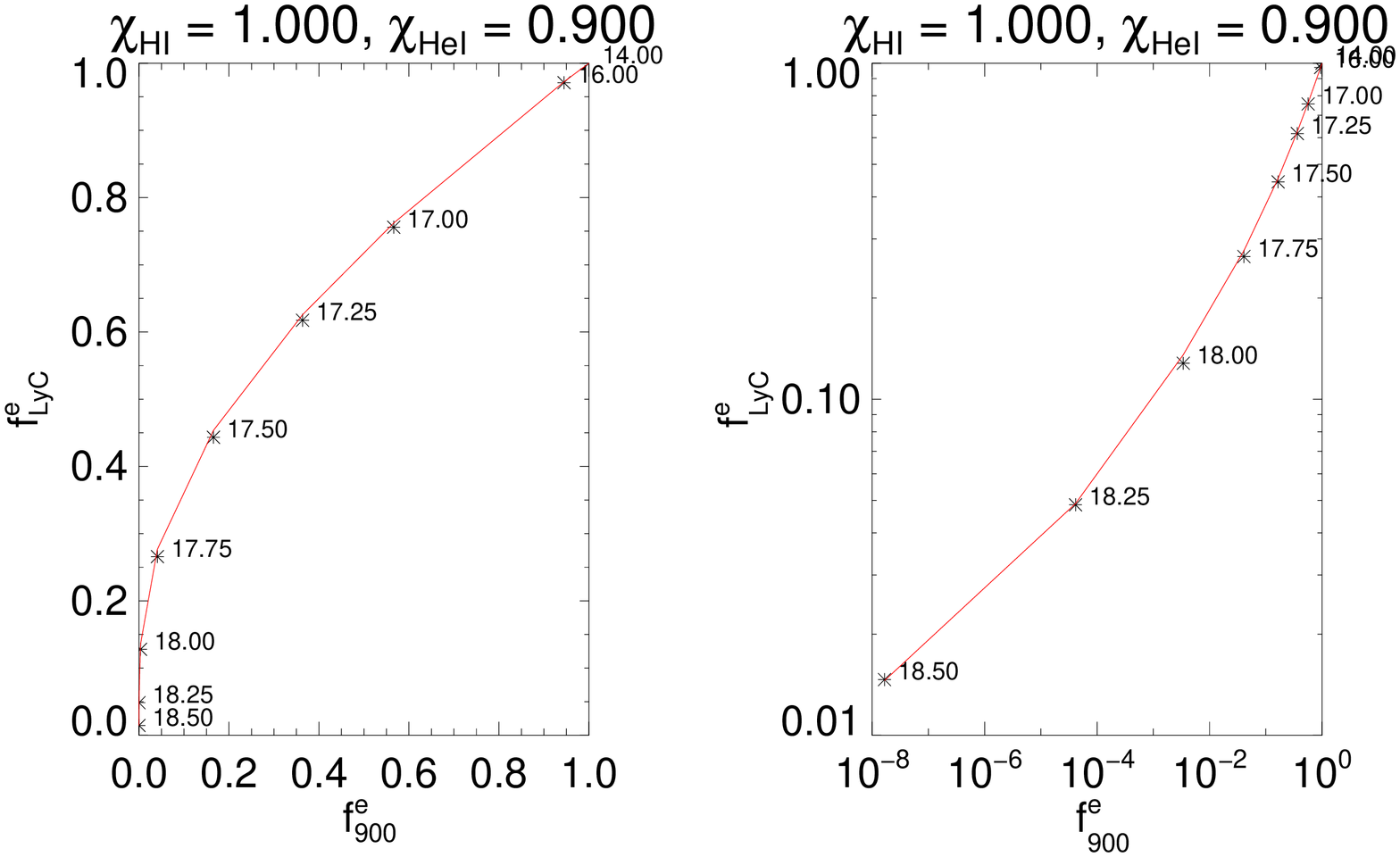}
\includegraphics[angle=90,width=0.47\textwidth]{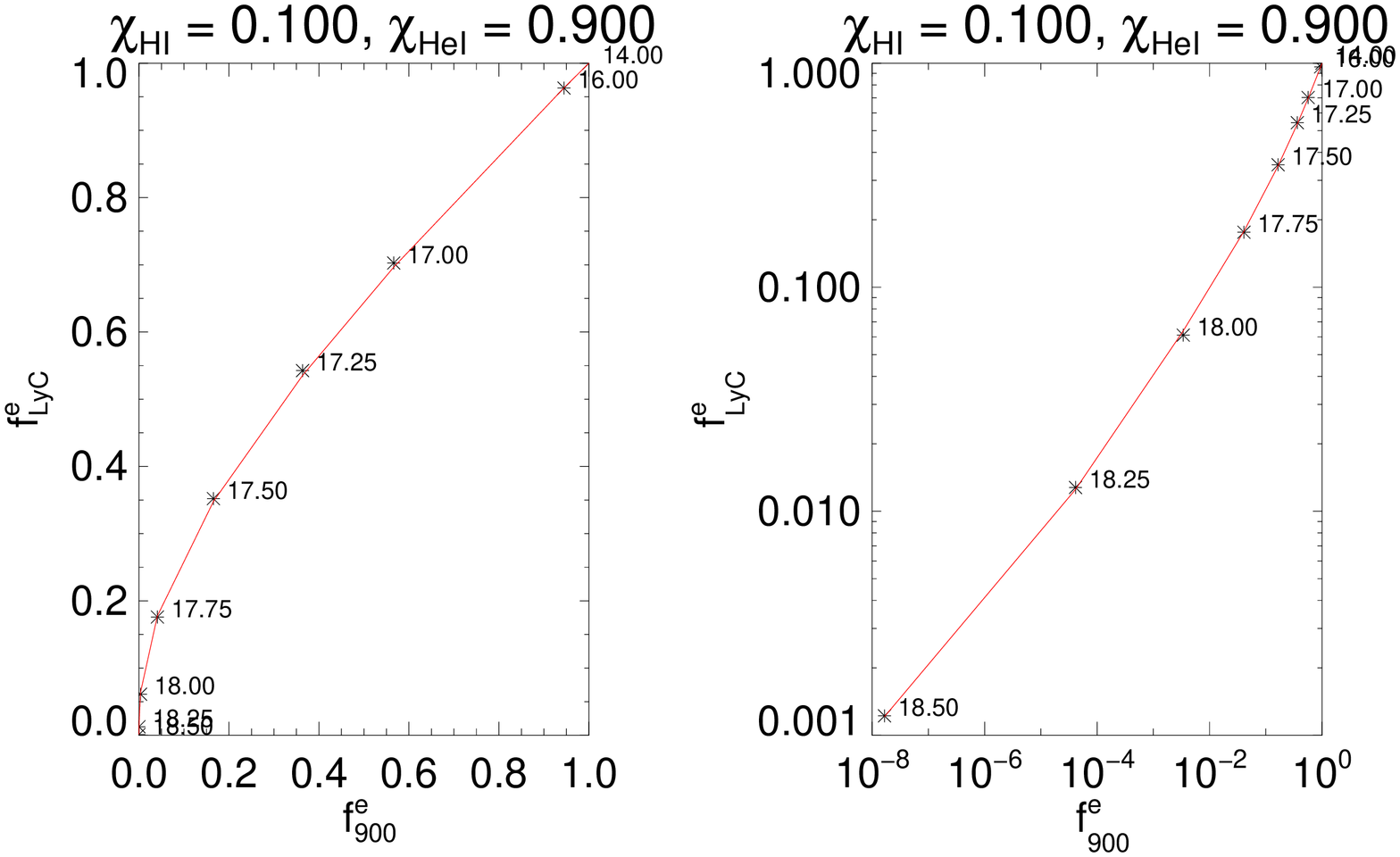}
\caption{Integrated fraction of escaping LyC photons, $f^e_{LyC}$ on ordinate, escape fraction at 900 $f^e_{900}$on abscissa.  Top--bottom order for  ($\chi_{HI}$, $\chi_{HeI}$) as in Figure~\ref{fig3}.  Left linear--linear.  Right log--log.    The relationship is decidedly non-linear.  Top to bottom shows varying degrees of significant escape even when $f^e_{900}$ has become black. \label{fig5}}
\end{figure}

In Figure~\ref{fig4} we show the effect of attenuating the SB99 model with the transmission functions depicted in Figure~\ref{fig3}.  The shape of the continuum emitted below the Lyman edge, and hence the number of ionizing photons that ultimately escape from a star-forming region, is linked to the ionization state of hydrogen and helium in the surrounding column of gas.  There is very little intrinsic stellar flux emitted below the \ion{He}{2} edge at 228 \AA, so the inclusion of \ion{He}{2} has little effect on the overall total escape fraction.   However, it would become important for the harder SEDs typical of quasars and AGN.  

%The top case where ($\chi_{HI}$, $\chi_{HeI}$) = (1.0, 0.01) is equivalent to neglecting helium altogether (and is likely unphysical as it represents the case strongly ionized helium and totally neutral hydrogen).  It shows a smooth recovery in flux toward shortward of 912$^-$.   

The top panel has ($\chi_{HI}$, $\chi_{HeI}$) = (1.0, 0.9), where we have assumed that helium is slightly ionized for the purpose of showing the location of the \ion{He}{2} Lyman series.  This model shows signs of a slight break shortward of the \ion{He}{1} ionization edge at 504 \AA, but overall it exhibits a relatively smooth Lyman drop-in with decreasing wavelength.  In the bottom panel ($\chi_{HI}$, $\chi_{HeI}$) = (0.1, 0.9),  so the break shortward of 504 \AA\ is strengthened by the lower \ion{H}{1} abundance with respect to \ion{He}{1}, causing a Lyman "double drop-in" to appear in the EUV region.  The strength of the \ion{He}{1} break with respect to the \ion{H}{1} break becomes more enhanced as $\chi_{HI}$ decreases.

In each figure we have tabulated logarithms of the \ion{H}{1} column density, the rate of ionizing photon production, the escape fraction integrated over the entire EUV band, $f^e_{LyC}$, and the escape fraction of ionizing photons in the narrow 20 \AA\ wide region shortward of the Lyman edge, $f^e_{900}$.  The table entries are color-coded to the spectra.  

In Figure~\ref{fig5} we show that the relationship between $f^e_{LyC}$ and $f^e_{900}$, (logarithmic scales left and liner scales right) is decidedly nonlinear.    We have overplotted an interpolation formula of the form

\begin{equation}
f^e_{LyC}  =\frac{(f^e_{900})^{\zeta_1} + (f^e_{900})^{\zeta_2} + \epsilon}{(2+\epsilon)},
\label{eq8}
\end{equation}

\noindent where ($\zeta_1, \zeta_2$, and $\epsilon$) = (0.25, 0.75, 0.018) and (0.37, 0.95, 0.0011), for ($\chi_{HI}$, $\chi_{HeI}$) = (1.0, 0.9) and ($\chi_{HI}$, $\chi_{HeI}$) = (0.1, 0.9), respectively.

In all cases a column density $\log{N_{HI}(cm^{-2})}$ = 18 produces essentially zero flux at the edge, yielding an edge escape fraction of $f^e_{900}$= 0.003, yet the integrated escape fraction is 13 and 6\%, for the two cases, respectively.  Using the interpolation formulae, we find that for the Lyman edge escape fractions used in Figure~\ref{fig2}, where $f^e_{900}$ = (0.01, 0.02, 0.04, 0.08, 0.16, 0.32, 1.00),  the corresponding integrated escape fraction is $f^e_{LyC}$ = (0.18, 0.22, 0.27, 0.35, 0.45, 0.59, 1.00)  and (0.10, 0.13, 0.18, 0.24, 0.34, 0.50, 1.00) for the two cases, respectively.  

We conclude that measurements of the Lyman edge escape fraction, $f^e_{900}$, only provide, at best, lower limits to the the true integrated fraction of escaping ionizing photons, and generally offer poor representations of the total number of ionizing photons that escape.    The tables in Figures~\ref{fig4} show that LBGs may emit significant amounts of LyC radiation, at the level of 5 to 1\%, even if the optical depth at the edge is $\approx$ 10 ( $\log{N_{HI}(cm^{-2})}$ = 18.25).   By way of example, we note that \citet{Izotov:2016a} found an edge escape fraction $f^e_{900}$ = 0.08.  Using Eq.~\ref{eq8}, we derived an integrated escape fraction of $f^e_{LyC}$  = 0.35 or 0.25 for our two cases, respectively.  This suggests that the ``f-escape'' problem might not be as bad as it seems.  The only definitive way to rule this out is through deep UV observations in the redshifted rest frame of the EUV in an attempt to observe the LyC drop-in. 

In the next section we address the observability of of the Lyman drop-ins in the face of the steadily increasing mean optical depth of the IGM with increasing redshift.

\section{Redshifted models, including mean IGM transmission  } \label{igmatten}

\begin{figure}
\includegraphics[width=0.5\textwidth]{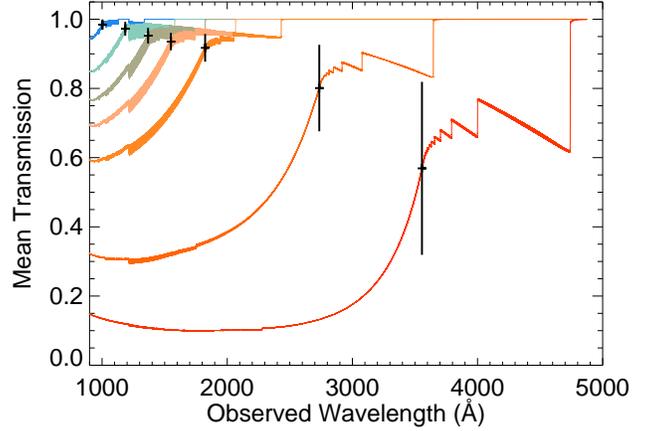}
\caption{Mean IGM transmission functions for redshifts $z$ = (0.1, 0.3, 0.5, 0.7, 1.0, 2.0, 2.9).  Computed from the ($\chi_{HI}$, $\chi_{HeI}$) = (1.0, 0.9) model.  The vertical bar indicates the level of expected variation at the edge as found in Monte Carlo study of IGM absorbers distributed over a range of column densities from LAF to DLA with a piecewise break at the LL transition \citet{Inoue:2008}.   \label{fig6}}
\end{figure}

\begin{figure*}
\includegraphics*[width=\textwidth,clip=false]{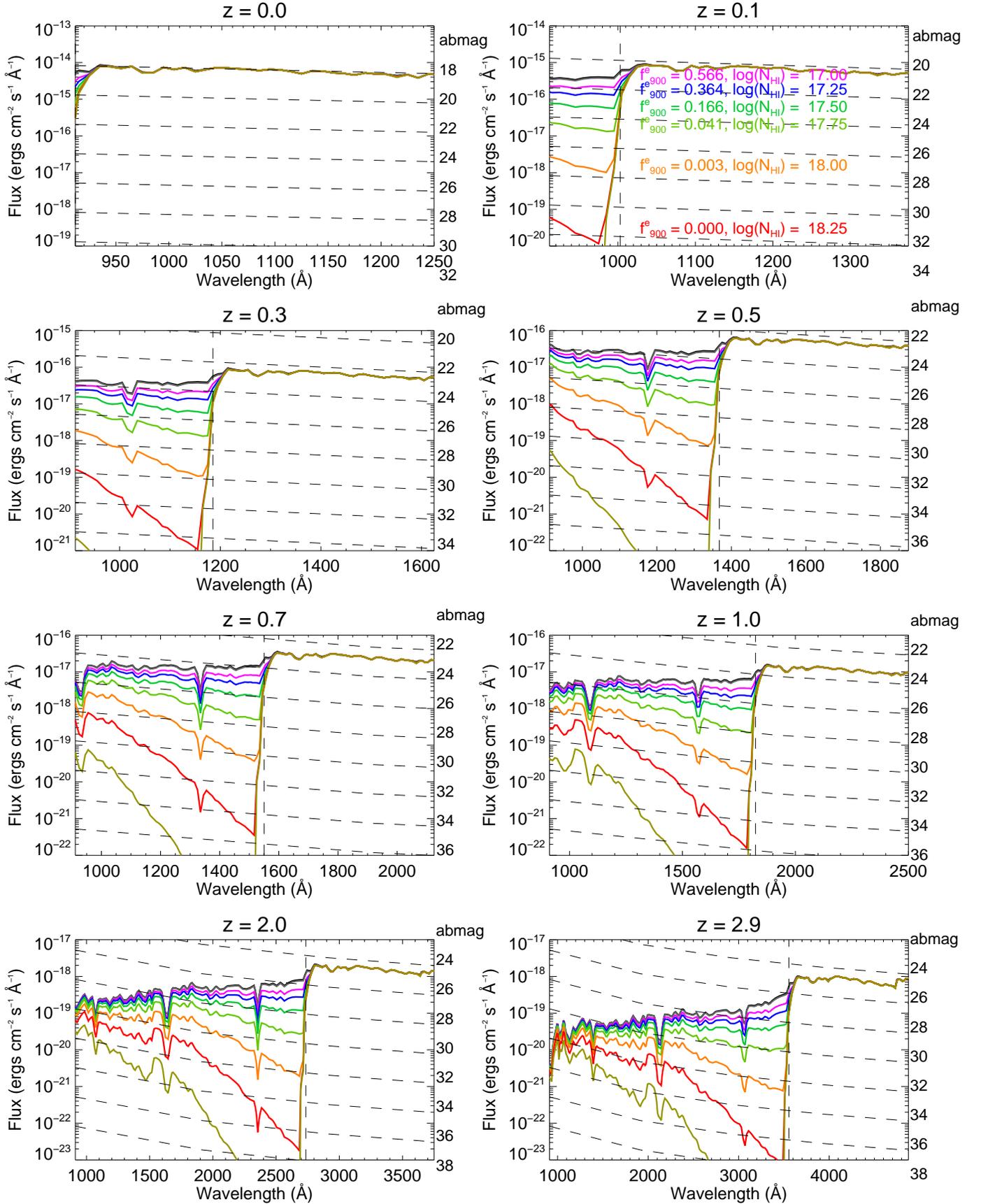}
\caption{Redshifted  ($\chi_{HI}$, $\chi_{HeI}$) = (1.0, 0.9) model with logarithmic scaling, showing LyC drop-in towards shorter wavelengths. Contours of constant abmag appear as dashed lines.  Column density colors as in Figure~\ref{fig4}.  The edge escape fractions are $f^e_{900}$ (\added{0.000,} 0.000, 0.003, 0.041, 0.166, 0.364, 0.566, 0.945)   for (\added{olive,} red, orange, light green, green, blue, violet, and grey) respectively.  \added{The $z$ = 0.1 panel, top-right, shows the  $f^e_{900}$ escape fractions and associated column densities.} \label{fig7}}
\end{figure*}

\begin{figure*}
\includegraphics*[width=\textwidth,clip=false]{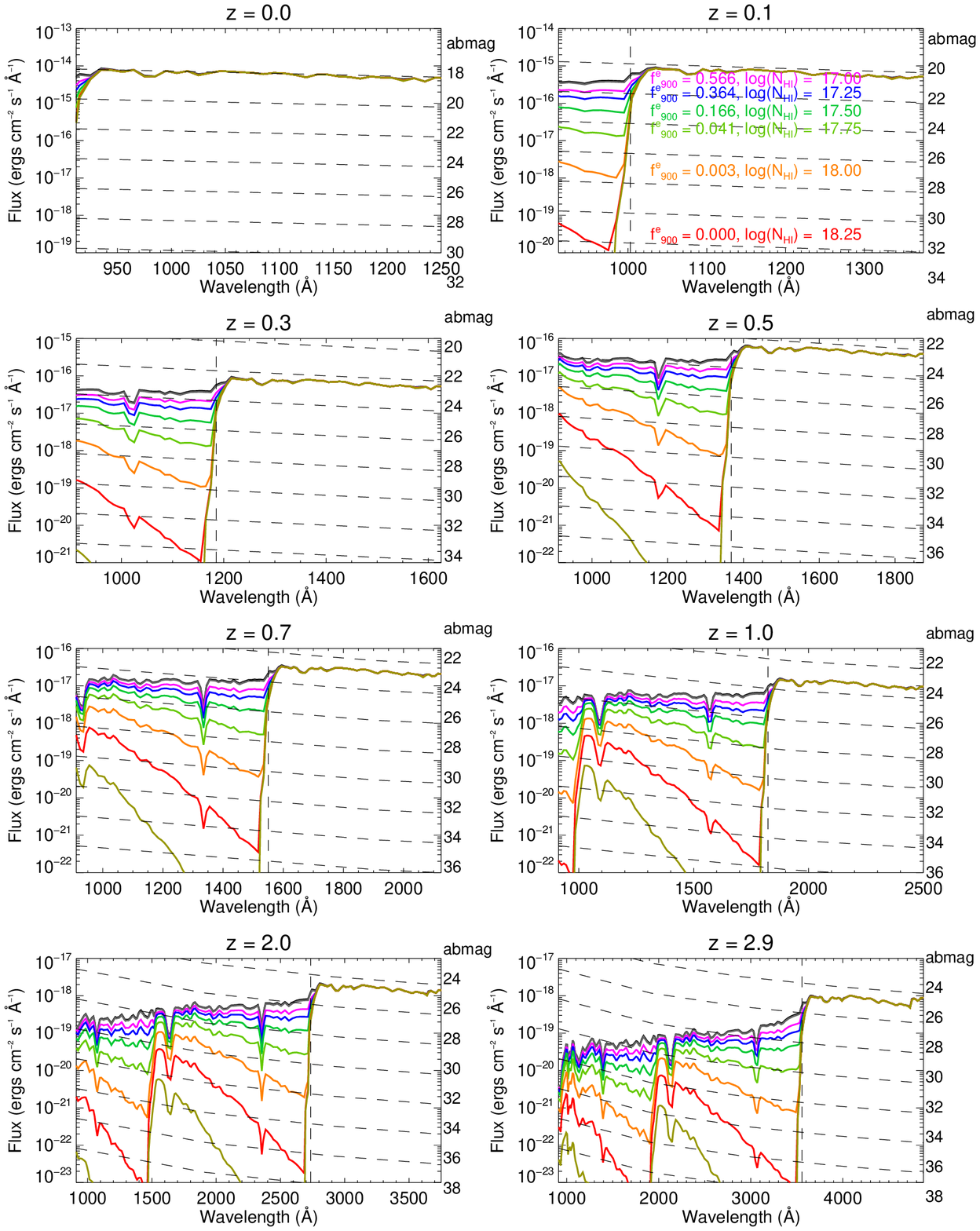}
\caption{Redshifted  ($\chi_{HI}$, $\chi_{HeI}$) = (0.1, 0.9) model with logarithmic scaling.  The $z$ = \replaced{1}{2.0} and \replaced{2}{2.9} models show the pronounced double drop-in structure is severely depressed, but the structure survives albeit at a difficult to detect level.  Colors and edge escape fractions as in Figure~\ref{fig7}.  \label{fig8}}
\end{figure*}

We have thus far not included attenuation of the escaping LyC due to resonant scattering and photoelectric absorption (photoionization) from discrete collections of intervening clouds in the IGM.   Here we use the rest frame LyC transmission models described in \S~\ref{LyCTrans} to compute the mean transmission of the IGM as a function of the fiducial redshifts in Table~\ref{t1}. 

%Here we use the distribution function from \citet{Madau:1995},

%\begin{equation}
%\left(\frac{\partial^{2}n}{\partial{N_{HI}}\partial{z}}\right)_{i,j}= \left \{ 
%\begin{array}{ll} 
%2.3 \times 10^{7}N_j^{-1.5}(1+z_i)^{2.46} & (12.3 < \log{(N_j)} < 17.2),\\
%1.9 \times 10^8N_j^{-1.5}(1+z_i)^{0.68} & (17.2 < \log{(N_j)} <   20 ),\\
%\end{array} \right.
%\end{equation}

%\noindent which  does not include contributions from the DLA regime as does the slightly more complicated form offered by \citet{Inoue:2014}.  However, \citet{Inoue:2014} show that the resulting mean transmission using the \citet{Madau:1995} distribution is lower than theirs.

\begin{figure*}
\includegraphics*[angle=90,width=\textwidth]{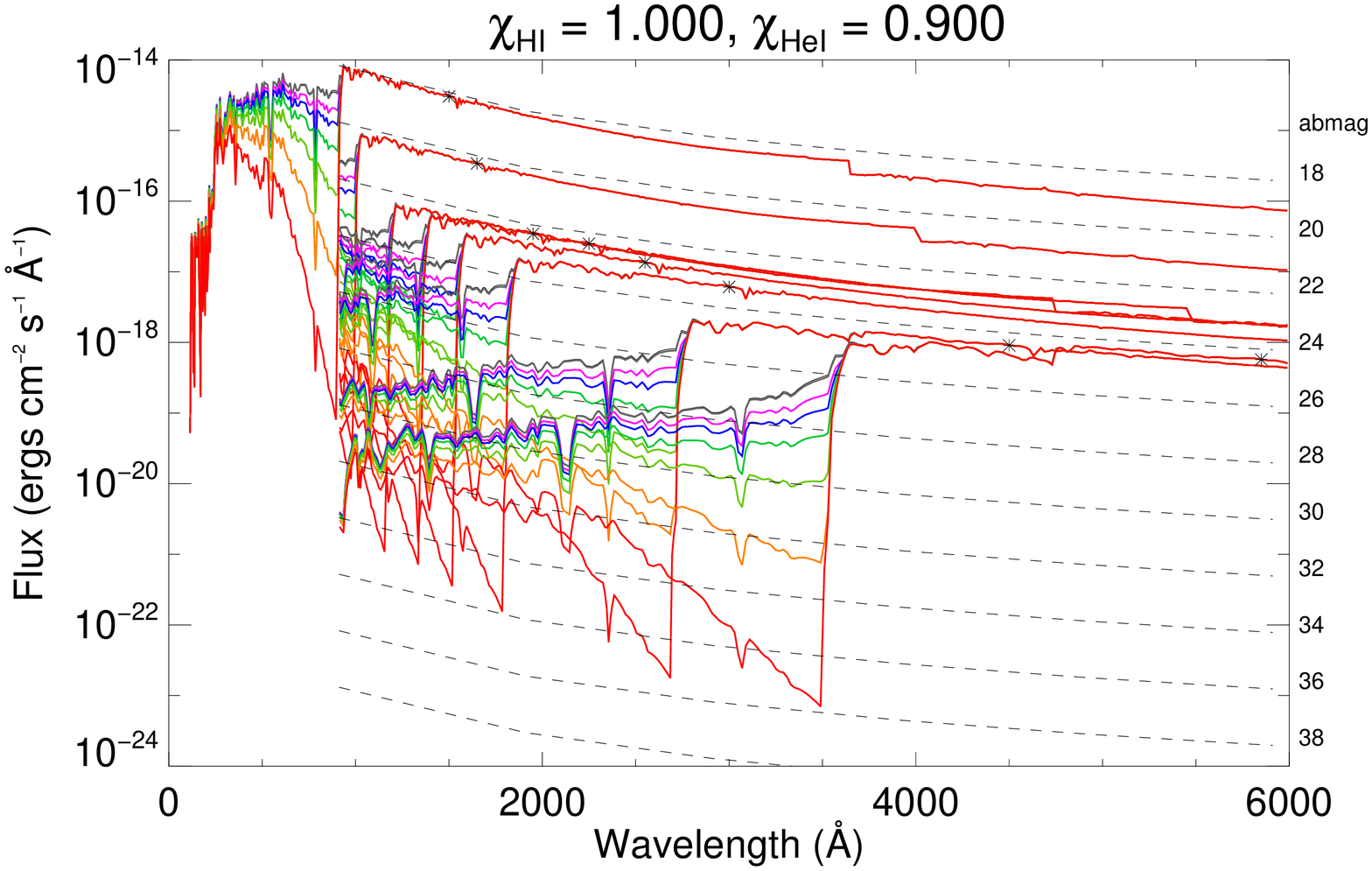}
\includegraphics*[angle=90,width=\textwidth]{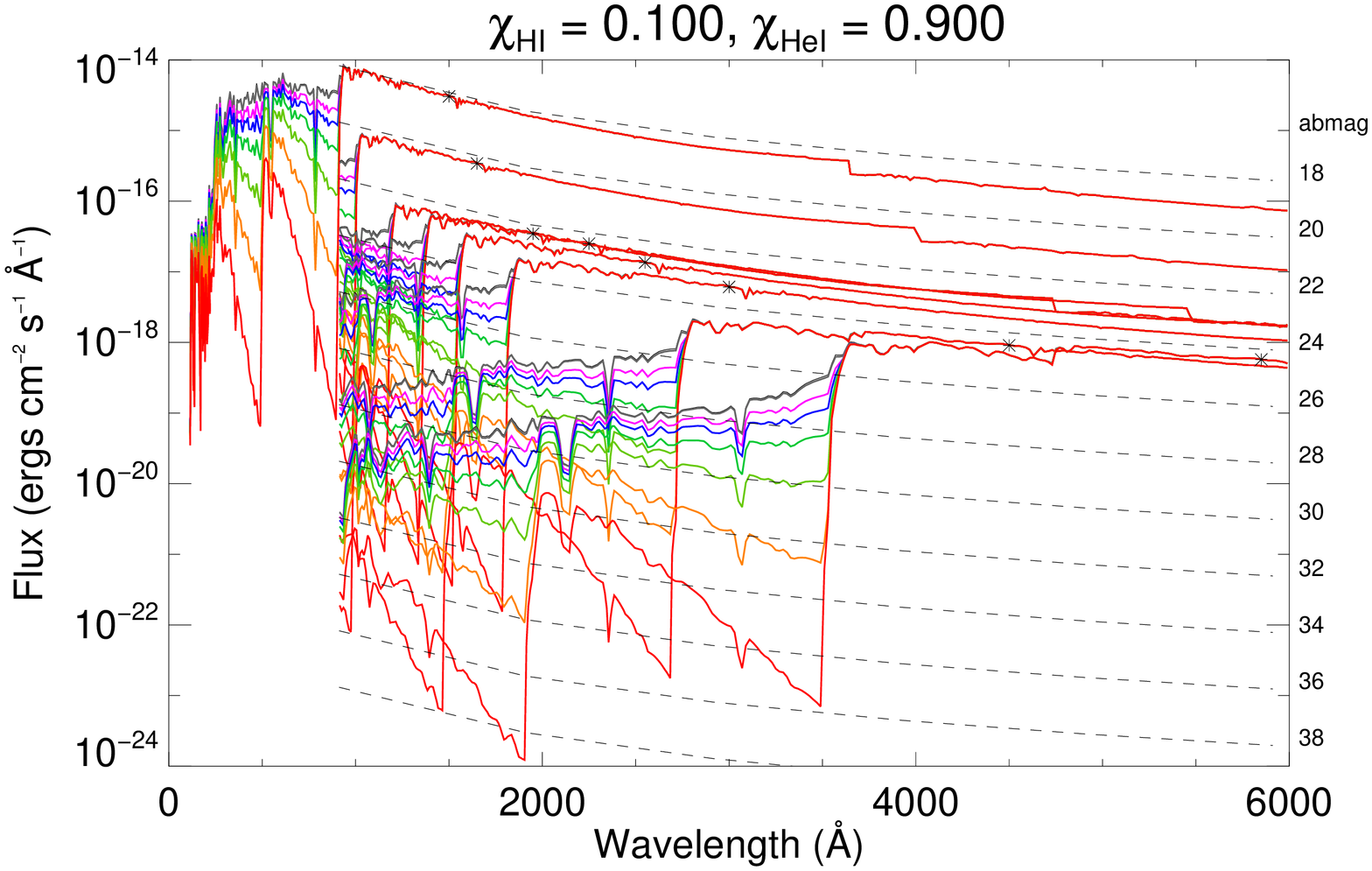}
\caption{\added{Attenuated SB99 models overplotted for the fiducial redshifts ($z$ = 0, 0.1, 0.3, 0.5, 0.7, 1.0, 2.0, 2.9).  The z = 0 models extend to 91 \AA; all others cut off at 911.8 \AA. }  Column density colors as in Figure~\ref{fig4}, but the lowest column 18.5 is omitted.  Asterisks mark (1+$z$)1500 \AA .  Contours of constant abmag appear as dashed lines.  Top --  ($\chi_{HI}$, $\chi_{HeI}$) = (1.0, 0.9) model with logarithmic scaling.  Bottom -- ($\chi_{HI}$, $\chi_{HeI}$) = (1.0, 0.9).      \label{fig9}}
\end{figure*}

The mean transmission function is derived from a mean optical depth \citep{Paresce:1980, Madau:1995, Inoue:2014}, expressed here as a numeric integral

\begin{equation}
<\tau(\lambda)> = \sum_{i=0}^{L}\sum_{j=0}^{M} \left(\frac{\partial^{2}n}{\partial{N_{HI}}\partial{z}}\right)_{i,j} (1-e^{-\tau(\lambda_{o})} )\Delta z_i  \Delta N_j . 
\end{equation}

\noindent Here the observed wavelength is related to the rest frame wavelength by $\lambda_{o} = \lambda_{r}(1+z_i)$.  The optical depth is $\tau(\lambda_{o}) = \tau_{HI}(\lambda_{o}) +\tau_{HeI}(\lambda_{o}) +\tau_{HeII}(\lambda_{o})$.  The redshift range from 0 $ \le z_i \le z_{L}$ is in fixed increments of $\Delta z_{i}$ =0.0005.  The logarithm of the column densities range from  12.3 $ \le \log({N_j}) \le$ 22 in variable increments of $\Delta N_{j} = N_{j+1/2} - N_{j-1/2}$ with 0 $ \le j \le M$ .  

The differential distribution of the number of discrete absorbers with respect to  \ion{H}{1}  column density and redshift, $\frac{\partial^{2}n}{\partial{N_{HI}}\partial{z}}$, is typically characterized by a piecewise continuous function in redshift and column density \citep[c.f.][]{Madau:1995, Inoue:2014}.   There are three basic \ion{H}{1} absorption regimes: the \lya\ forest regime (LAF), $  \log{N_{HI}(cm^{-2})} \lesssim $ 17; the Lyman limit (LL) regime,  17 $\lesssim \log{N_{HI}(cm^{-2})} \lesssim$ 20, and the damped \lya\ (DLA) regime,  $ 20 \lesssim \log{N_{HI}(cm^{-2})} $.   

We employ the $\frac{\partial^{2}n}{\partial{N_{HI}}\partial{z}}$  described by \citet{Inoue:2014}, which uses a "Schechter-like" distribution function for the column density multiplied by a piecewise-continuous multiple power law distribution to describe the evolution in redshift.    The distribution function is broken down into the sum of two different products, which in combination reasonably agrees with the evolution of \ion{H}{1} absorption systems from the LLA  to  DLA regimes as detailed in the literature \citep{Weymann:1998, Prochaska:2005, Prochaska:2010, Ribaudo:2011, Noterdaeme:2012, Fumagalli:2013, Kim:2013, O'Meara:2013, Prochaska:2014}.   We note that the \citet{Inoue:2014} distribution function produces a mean transmission that is generally less aggressive than that produced by the distribution function used by \citet{Madau:1995}, even though it does not include a contribution from DLAs.

The mean IGM transmission functions in the observers frame, $T_{IGM}(\lambda_o) = \exp{<-\tau(\lambda)>}$,  are plotted in Figure~\ref{fig6} for $z$ = (0.1, 0.3, 0.5, 0.7, 1.0, 2.0, 2.9).  These transmission functions were computed from the ($\chi_{HI}$, $\chi_{HeI}$) = (1.0, 0.9) model.  We emphasize that they are just mean relationships.  Significant stochastic deviations are expected along any given line of sight.   We give an indication of the level of such variations at the Lyman edge by overplotting the 68\% deviations found by \citet[][their Figure 8]{Inoue:2008} in a Monte Carlo simulation of IGM absorbers distributed over a range of column densities from LAF to DLA with a piecewise break at the LL transition.  The appearance, or absence, of an LL system in the Monte Carlo simulation drives much of the expected transmission stochasticity.   We see that while the mean attenuation is significant at $z$ =2.9, reaching a lower trough of $\approx$ 15\% at 2000 \AA, it is by no means complete.  The transmission at the Lyman edge  for each redshift is  $T(z)$ = (0.98, 0.97, 0.95, 0.94, 0.92,      0.80, 0.57), which translates to a decrease in abmag at the respective edges of $\delta m^*_{(1+z)900}$ = (0.02, 0.03, 0.05, 0.07, 0.09, 0.24, 0.61).  
 
In Figures~\ref{fig7}, \ref{fig8} and \ref{fig9} we show the result of applying $T_{IGM}(\lambda_o)$ to the ($\chi_{HI}$, $\chi_{HeI}$) = (1.0, 0.9) and ($\chi_{HI}$, $\chi_{HeI}$) = (0.1, 0.9) luminosity density models respectively.  \added{Following \citet{Oke:1983},} using abmag = -2.5$\log{F_{\nu}}$-48.6 and substituting $F_{\nu} = \frac{\lambda^2}{c}F_{\lambda}$, we determined the flux, $F_{(1+z)1500}$  in the observer's frame from \deleted{the characteristic abmag}  $m^*_{(1+z)1500}$ \deleted{of Eq.~\ref{eq5}} in Table~\ref{t1}, such that
\begin{equation}
\log(F_{(1+z)1500}) = \frac{m^*_{(1+z)1500} + 5\log{[(1+z)1500]} +2.408}{-2.5},
\end{equation}
\explain{This formula has been corrected.  Originally there was a 4 in front of the log when it should have been a 5.  The zero point has been changed from 2.406 to 2.408 = 48.6 - log c, where c is the speed of light expressed in \AA\ s$^{-1}$.}

\noindent and found a scale factor, $M_7$, matching the redshifted SB99 luminosity density at luminosity distance $d_{l}(z)$ to the observed SED $F_{(1+z)1500}$ such that

\begin{equation}
F_{\lambda_o} = M_7\left(\frac{L_{(1+z)\lambda_r} }{4 \pi d_{l}^{2}(z) (1+z)}\right).
\end{equation}

\noindent Here the luminosity distance $d_{l}(z)$ is in cm \added{and $M_7$ = $\frac{M^{\star}_g}{(10^7 M_{\odot})}$, where $M^{\star}_g$ is the galaxy stellar mass and $10^7 M_{\odot}$ is the mass of the SB99 model with its SFR of 1 $M_{\odot}$ yr$^{-1}$ and age of 10$^7$ years.  We find} $M_7 = $ [0.8, 1.1, 3.2, 4.4, 5.6, 6.9, 13.7]  for the fiducial redshifts in Table~\ref{t1}.   The $z = 0 $ model is fixed to $m_{1500}$ = 18.0, with $M_7 =1 $ at a distance of $d_{l}$ = 177 Mpc.  The mean \added{IGM} transmission function was applied as a uniform screen \replaced{after redshifting and scaling}{to $F_{\lambda_o}$.  The CGM transmission is applied to $L_{\lambda_r}$ in the rest frame prior to redshifting.}    Lines of constant abmag are overplotted on the figures.

\subsection{LyC Detection Requirements \label{detreq}}

The results of the previous section show that detection and quantification of the fraction of LyC flux escaping from star-forming galaxies out to $z \sim$ 3 will be a formidable, but not  insurmountable, challenge for future space observatories.  Moreover, they are consistent with the paucity of detections to date.   

\subsubsection{Current Capabilities}
Examining the $z =$ 0.1 models, we see that very little of the LyC region peeks out below $\sim$ 1000 \AA.   The Cosmic Origins Spectrograph (COS) on the {\it Hubble Space Telescope} (\hst)  has limited sensitivity in this bandpass ($A_{eff} \sim$ 10 cm$^{-2}$), with a background equivalent flux (BEF) of $\sim$ 10$^{-15}$ \flux\ $\sim$  20 abmag \citep{McCandliss:2010, Redwine:2016}.  Galaxies brighter than this are in the region of the exponential falloff in the luminosity function, so they become increasingly rare with increasing luminosity.  Characteristic galaxies with edge escape fractions $f^e_{900} < 0.4$, will have an attenuated edge flux more than an order of magnitude lower than the COS BEF. Consequently, they will be background limited, requiring extraordinarily long integration times to detect with confidence.  We can expect to detect only a handful of such objects over the lifetime of COS.  The three detections by \citet{Leitherer:2016} at $z \approx$ 0.04 had edge fluxes $\sim$ 1 to 2 $\times$ 10$^{-15}$ \flux. 

The situation is somewhat more favorable at $z$ = 0.3, 0.5, and 0.7, where the  $m^*_{(1+z)1500}$ are quite similar  ($m^*_{1950}$, $m^*_{2250}$, $m^*_{2550}$ = 22.3, 22.4, 22.7)  and the LyC region extends shortward of  $\sim$ 1200, 1350, and 1550 \AA, respectively, into a wavelength region where the COS effective area $A_{eff} >$ 1000 cm$^{-2}$.  Although the $m^*_{(1+z)1500}$ for galaxies at these redshifts are nearly an order of magnitude fainter than at $z$ = 0.1, the COS BEF is much lower than at $z =$ 0.1 ($\sim$ 10$^{-17}$ to 10$^{-18}$ \flux\ $\sim$ 25 to 27.5 abmag -- depending on orbital attitude and solar activity),  thus allowing for more efficient observing programs to be constructed.  Still, the background limit is reached for characteristic galaxies at $f^e_{900} \approx$ 0.04 to 0.16.  The depth of such observations will be limited and robust statistical samples difficult to come by, especially for $z$ = 0.5 and 0.7, where the number density is nearly 4 times lower.   Nevertheless, the spectral baseline is relatively longer, which along with the higher effective area favors the detection of LyC drop-ins toward these higher redshifts.

For galaxies at $z \approx$ 1 the detection of escaping LyC with COS becomes even more difficult, as the unattenuated LyC region is of order the BEF.  This was also the case for LyC leak photometric searches conducted \replaced{by}{using} \galex.  The FUV ($\lambda_{eff}$ = 1528 \AA, $\Delta\lambda$ =1344 - 1786)  and  NUV ($\lambda_{eff}$ = 2271 \AA, $\Delta\lambda$ =1771 - 2831 channels on \galex\ straddle the break for objects at a redshift of $z \approx$ 1.1.  The \galex\  5$\sigma$ flux limit was $\sim$25 to 24 abmag for 30 ks integration in the FUV and NUV, respectively \citep{Morrissey:2005}.  In the NUV  a $z$ = 1  characteristic galaxy  is $\sim$23 abmag, while the unattenuated FUV SED  ($f^e_{900}$ =1) has $\sim$25 abmag at 1530 \AA; this is too faint for robust detections at near unity escape fraction.  These limits are consistent with the \citet{Cowie:2009} analysis of deep \galex\ exposures of the GOODS-N field with spectroscopic redshifts, where  they  found no objects with FUV $<$ 25 for redshifts of $z > 0.6$.   We conclude that the limiting flux of \galex\ was not low enough to adequately survey \lyc\ leakage at a redshift of $z \ga$ 1.

Observations by \citet{Siana:2007} using the Space Telescope Imaging Spectrograph (STIS) and Advanced Camera for Surveys Solar Blind Channel (ACS/SBC) found no detections in deep far-UV observations around 1600 \AA\  of 21  HUDF objects with known spectroscopic redshifts 1.1 $< z < $ 1.5.  The $3\sigma$ limits in the ACS/SBC F150LP channel were estimated to be $\approx$28 abmag.  Their median abmag at 1500(1+1.3) = 3450 \AA\ was 24.5, which is about 1 mag fainter than our characteristic magnitude at $z$ = 1.  The non-detection to a 28 abmag limit at observer frame 1600 \AA\ implies a $\log{N_{HI}(cm^{-2})} \ga$ 17.75, yielding a $f^e_{900}$ $\sim$ 0.04  (an integrated 0.17 $< f^e_{LyC} <$  0.26).  These values compare fairly well to their stacked limit of $f_{esc,rel} <$ 0.08 evaluated at 700 \AA\ in the rest frame.

\added{\citet{Naidu:2016} analyzed} Hubble Deep UV (HDUV) imaging of the GOODS-north and -south fields, using the Wide-Field Camera 3 (WFC3) filters F275W and F336W and redshifts supplied by 3D-\hst\ Grism,   \deleted{were used} to  explore the redshift range $z \sim$ 2--3.  The 5$\sigma$ detection limit was $\sim$ 28 abmag.  The candidate leakers were estimated to have $f_{esc} >$ 0.6.  This  result is consistent with our $z $ = 2 and 2.9 calculations, showing that a characteristic galaxy at these redshifts with an abmag of 28 in the LyC will have $f^e _{900}>$ 0.4 at $z$ = 2 and $f^e _{900} >$ 0.6 at $z$ = 3.

\begin{figure}
\includegraphics[width=.5\textwidth]{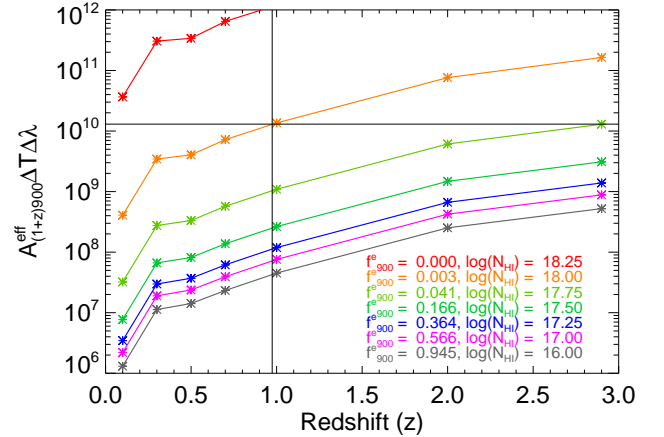}
\caption{ The product of effective area $A_{eff}(\lambda)$, observing time $\Delta T$, and bandwidth $\Delta \lambda$
required to detect flux from a characteristic galaxy, attenuated by a $\log{N_{HI}(cm^{-2})}$ = [18.25, 18.00, 17.75, 17.50, 17.25, 17.00, 16.00], in the wavelength region just over the Lyman edge, with a signal to noise of 5.  The corresponding edge escape fractions are $f^e_{900}$ = [0.000, 0.003, 0.041, 0.166, 0.364, 0.566, 0.945] in color code of [red, orange, light green, green, blue, violet, grey].  The example given in the text (Eq.~\ref{eq12}) is marked with black horizontal and vertical lines.
  \label{fig10}}
\end{figure}

\subsubsection{Future Capability Requirements \label{future}}

It is clear from Figure~\ref{fig2}  that if we are to achieve the goal of statistically significant determination of LyC luminosity function evolution wherein we probe down to $f^e_{900} \approx$  0.01 out to a redshift of 3, then we need to reach an abmag  from 25 to  30 for the characteristic objects in the redshift range from 0.1 $< z < $ 2.9 .  This corresponds to a flux range of  $F_{(1+z)900} \sim$ 10$^{-17}$ to 10$^{-20}$ \flux.  The luminosity function spans a range of $\approx \pm$ 2.5 mag about the characteristic value, so the flux range will need to be another factor of 10 lower to obtain adequate numbers of  low-luminosity objects.

By way of example, \replaced{let us}{if we} assume a limiting flux requirement of $F^{lim}_{1800}$ =2 $\times$ 10$^{-20}$ \flux\ at a 1800 \AA\ as a representative goal for $\sigma$ = 5  of edge detection with a $f^e_{900} \ga$ 0.003 for faint-end objects at $z \approx$ 1, then we arrive at the following criteria for the product of effective area ($A^{eff}_{1800}$, observing time $\Delta T$, and $\Delta \lambda$ spectral bandwidth, assuming negligible background):
\begin{equation}
 A^{eff}_{1800}~\Delta T~\Delta \lambda = \frac{\sigma^2}{F_{1800}/E_{1800}} = 1.3 \times 10^{10} \cmtwosa,
 \label{eq12} \end{equation}
 
\noindent where $E_{1800}$ is the photon energy at 1800 \AA.  This formula assumes that there is negligible background.   It may be used for estimating effective area, observing time, and bandwidth requirements for photometric and spectroscopic applications.   In Figure~\ref{fig10} we show the area-time-bandwidth product Lyman edge detection requirement for $\log{N_{HI}(cm^{-2})}$ = [18.25, 18.00, 17.75, 17.50, 17.25, 17.00, 16.00] as a function of redshift.

\subsubsection{Target Sample Goals }

\added{Figure~\ref{fig1} shows that} there are $\sim$ 300 objects per unit magnitude per square degree at $z =$ 0.1 at the faint-end.  By 0.3 $z =$ 0.3 the faint end areal density are $\sim$ 3000 objects per unit magnitude per square degree, being flat out to $z =$ 1.  For $z =$ 2 the areal density at the faint end rises to $\sim$ 50,000 objects per unit magnitude per square degree.    

The construction of high-fidelity luminosity functions places a requirement on sample size, where 25 objects per luminosity bin per redshift interval will yield an approximate rms deviation of $\sim$ 20\% for each point. Luminosity range should be ~ 2 - 3 orders of magnitude with $\sim$ 20 to 40 bins covering 10 redshift intervals. Redshifts are required for each object. This suggests an observing program with 500--1000 objects per redshift interval, yielding 5000--10,000 objects in total.  

The development of a wide-field multi-object spectroscopic (MOS) capability and complementary photometric observations will be essential to providing these samples.  Angular coverage requirements will be driven by low-redshift objects, which have lower areal number densities than higher-redshift objects, while observing time requirements will be driven by the faintest objects.

\subsection{Observing Strategies}

Spectroscopy and photometry provide complementary approaches to this problem. Spectroscopy offers an opportunity to examine in detail the variation of the flux escaping at wavelengths below the Lyman edge along with the compositional properties of the objects.  Photometry can go deeper and offer higher spatial resolution at the expense of spectral information. Spectroscopy provides an avenue for training photometry.

A comprehensive program will incorporate photometry into the search for candidates and the deep probe of the faint end of the LyC luminosity function, while low-resolution precision spectroscopy will be used to characterize the shape of the LyC in a search for drop-ins, offering clues for the total ionizing photon budget and potentially dust attenuation in a completely unexplored bandpass.  Spectral multiplexing over several square arcminutes will be of great importance for acquiring statistically significant samples in a reasonable amount of observing time.

These observations should be complemented by medium- to high-resolution ($R \sim$ 10,000 - 50,000) spectroscopy longward of the Lyman edge,  providing a baseline assessment of intervening \ion{H}{1} and metallic absorption systems associated with the CGM and IGM.  Such observations, supported by spatially resolved imaging at the scale of 10 to 100 pc, can be used to account for the partial covering of emission regions, wherein the uniform screen assumption assumed here could be relaxed in favor of the informed modeling of foreground sources using measured superpositions of \ion{H}{1} columns and metallic absorption distributions.  This is especially important for determining the randomly distributed IGM attenuation prior to quantifying the actual escape from the CGM for the higher-$z$ objects.   Success at high $z$ is likely to rely on the identification of ``lucky sight-lines''  aided by gravitational lens and/or a several $\sigma$-low deviation from the average number of obscuring line of sight IGM absorption systems. \\

%The determination of \lyc\ escape from low redshift galaxies and investigating whether a correlated relationship exists, goes beyond the scope of this study.  However, the prospects for carrying it out have been bolstered by reports of successful detections at both low and intermediate redshift, although these, mostly low \fec, detections are not without contention \cite[c.f.][]{Borthakur:2014, Leitet:2013, Bergvall:2013, Nestor:2013, Iwata:2009, Grimes:2007, Bergvall:2006}. In the future, searches for \lyc\ from low redshift galaxies can be carried out by taking advantage of the recently appreciated Lyman edge limited sensitivity of COS \citep{McCandliss:2010}, or from newly proposed spacebased instruments, such as SubLymE \citep{Green:2014} or a FORTIS like instrument \citep{McCandliss:2008}. 

\section{Conclusion} \label{sumcon}

Measurements of $f^e_{LyC}$ at $z \la$ 3 across all types of star-forming galaxies are crucial to informing our understanding of how the universe came to be ionized.  UV observations, far and near, provide the most direct path to the spatially resolved detection of ionizing radiation, and the opportunity to characterize the physical state of galactic environments that favor LyC escape.

The primary purpose of this work is to provide estimates of the flux in the LyC from star-forming galaxies, as  functions of escape fraction and redshift.  \deleted{,  for the purpose of} \added{These estimates are useful for} quantifying science return,  \replaced{developing}{defining} technical requirements, and \added{developing} observing strategies for future large, medium, and small scale missions to be considered by the Astrophysics Decadal for 2020.  A model has been developed based on a uniform foreground screen of \ion{H}{1}, \ion{He}{1}, and \ion{He}{2}, attenuating a young continuous star-forming object \replaced{, having}{with}  fluxes at  (1+z)1500 \AA\ \replaced{set to }{scaled to reproduce} the characteristic apparent magnitudes of the \citet{Arnouts:2005} luminosity functions.  The uniform screen model has as free parameters the ionization fractions of H and He ($\chi_{HI}$, $\chi_{HeI}$), wherein the ratio of H to He is set to the cosmic abundance.  We also account for the mean attenuation by power law distributions of \lya\ forest, LL, and DLA systems associated with the IGM as a function of increasing redshift.

\added{The paucity of current detections to date are broadly consistent with the characteristic flux levels found here.  Sensitivity to an abmag $\approx$ 30 is required to detect an attenuated edge from a characteristic galaxy with $f^e_{900} \approx 0.003$, ($\log{N_{HI}(cm^{-2})} = 18.0$) out to $z \la$ 1.  For redshifts 1 $\la z \la$ 2.9, an abmag of 31 $\la m_{(1+z)900} \la$ 33 is required.}  

We \replaced{find}{caution} that the escape fraction measured just below the Lyman edge ($f^e_{900}$) at 911.8 \AA\ provides only a lower limit to the  fraction of LyC photons that escape as integrated over the entire EUV region ($f^e_{LyC}$).  For 18.5 $ > \log{N_{HI}(cm^{-2})} > $17.9, the Lyman edge can appear black (essentially zero) yet the integrated escape fraction can range from 1 to 20\%, depending on the choice of ionization fractions ($\chi_{HI}$, $\chi_{HeI}$).  

More accurate assessments of the integrated escape fraction could be made from spectroscopic observations that resolve the recovery of flux $\propto (\frac{\lambda}{\lambda_e})^3$; a phenomena that we have dubbed Lyman ``drop-ins''.  Observations of the LyC spectral shape could provide, in principle, the following: enhanced fidelity to the determination of the contribution of star-forming galaxies to the MIB; important observational insights into the  temperature of the CGM and IGM; and  constraints on the survivability and attenuation properties of dust grains with respect to the total gas content in LyC-leaking environments.

The uniform screen model adopted here is considerably simpler than the more realistic case of an undulating, highly discontinuous and chaotic ``unity \ion{H}{1} optical depth'' surface of the CGM surrounding a star-forming galaxy, which must have a $\tau$ low enough for LyC radiation to  escape.  The model should be thought of as a kind of ensemble average of the galaxy's partially covered star-forming regions, wherein much of the total star-formation may be buried by a more heavily attenuating ISM. 

We suggest a strategy of high spectral and spatial resolution observations in the region longward of the Lyman edge,  to  guide assessment of the line of sight geometry in coordination with a spectroscopic determination of the degree of partial covering, along with the identification of specific \lya\ forest, LL, and DLA systems that may contribute to the attenuation of objects toward higher redshift.  Such information will allow a relaxing of the simple uniform screen model in favor of more realistic radiative transfer models.

\deleted{The paucity of current detections to date are broadly consistent with the characteristic flux levels found here.  An abmag $\approx$ 30 is required to detect an attenuated edge from a characteristic galaxy with $f^e_{900} \approx 0.003$ out to $z \la$ 1.  At 1 $\la z \la$ 2 an abmag $\approx$ 32 is required.}  Achieving the goal of a statistically significant characterization of LyC luminosity function evolution out to $z \la$ 3 can be carried out most efficiently with a wide-field multi-object spectroscopic survey supported by photometric observations.  An instrument with an effective area, observing time, and spectral bandwidth product of 1.3 $\times$ 10$^{10}$ cm$^{2}$~s~\AA\ at 1800 \AA\ would be sufficient to carry out such a program to a redshift of $z =$ 1.

%This work has significant implications for quantifying the number of LyC photons that escape from star-forming galaxies and their potential to ionize the universe. It will also lead to important observational insights into the survivability of dust in highly ionized media.

\acknowledgments

We acknowledge useful conversations with Brian Fleming, Kevin France, Keith Redwine, Jason Tumlinson, and Harry Ferguson.  We thank the anonymous referee for keen suggestions that improved the readability and utility of this article.  Support for this work was provided by NASA to the Johns Hopkins University through APRA grants NNX08AM68G, NNX11AG54G, and NNX17AC26G.

\listofchanges

\begin{thebibliography}{}
\expandafter\ifx\csname natexlab\endcsname\relax\def\natexlab#1{#1}\fi

\bibitem[{{Arnouts} {et~al.}(2005){Arnouts}, {Schiminovich}, {Ilbert},
  {Tresse}, {Milliard}, {Treyer}, {Bardelli}, {Budavari}, {Wyder}, {Zucca}, {Le
  F{\`e}vre}, {Martin}, {Vettolani}, {Adami}, {Arnaboldi}, {Barlow}, {Bianchi},
  {Bolzonella}, {Bottini}, {Byun}, {Cappi}, {Charlot}, {Contini}, {Donas},
  {Forster}, {Foucaud}, {Franzetti}, {Friedman}, {Garilli}, {Gavignaud},
  {Guzzo}, {Heckman}, {Hoopes}, {Iovino}, {Jelinsky}, {Le Brun}, {Lee},
  {Maccagni}, {Madore}, {Malina}, {Marano}, {Marinoni}, {McCracken}, {Mazure},
  {Meneux}, {Merighi}, {Morrissey}, {Neff}, {Paltani}, {Pell{\`o}}, {Picat},
  {Pollo}, {Pozzetti}, {Radovich}, {Rich}, {Scaramella}, {Scodeggio},
  {Seibert}, {Siegmund}, {Small}, {Szalay}, {Welsh}, {Xu}, {Zamorani}, \&
  {Zanichelli}}]{Arnouts:2005}
{Arnouts}, S., {Schiminovich}, D., {Ilbert}, O., {et~al.} 2005, \apjl, 619, L43

\bibitem[{{Bahcall} \& {Sargent}(1967)}]{Bahcall:1967}
{Bahcall}, J.~N., \& {Sargent}, W.~L.~W. 1967, \apjl, 148, L65

\bibitem[{{Bechtold} {et~al.}(1987){Bechtold}, {Weymann}, {Lin}, \&
  {Malkan}}]{Bechtold:1987}
{Bechtold}, J., {Weymann}, R.~J., {Lin}, Z., \& {Malkan}, M.~A. 1987, \apj,
  315, 180

\bibitem[{{Benson} {et~al.}(2013){Benson}, {Venkatesan}, \&
  {Shull}}]{Benson:2013}
{Benson}, A., {Venkatesan}, A., \& {Shull}, J.~M. 2013, \apj, 770, 76

\bibitem[{{Bland-Hawthorn} \& {Maloney}(1999)}]{Bland-Hawthorn:1999}
{Bland-Hawthorn}, J., \& {Maloney}, P.~R. 1999, \apjl, 510, L33

\bibitem[{{Borthakur} {et~al.}(2014){Borthakur}, {Heckman}, {Leitherer}, \&
  {Overzier}}]{Borthakur:2014}
{Borthakur}, S., {Heckman}, T.~M., {Leitherer}, C., \& {Overzier}, R.~A. 2014,
  Science, 346, 216

\bibitem[{{Bouwens} {et~al.}(2015){Bouwens}, {Illingworth}, {Oesch}, {Caruana},
  {Holwerda}, {Smit}, \& {Wilkins}}]{Bouwens:2015}
{Bouwens}, R.~J., {Illingworth}, G.~D., {Oesch}, P.~A., {et~al.} 2015, \apj,
  811, 140

\bibitem[{{Cooke} {et~al.}(2014){Cooke}, {Ryan-Weber}, {Garel}, \&
  {D{\'{\i}}az}}]{Cooke:2014}
{Cooke}, J., {Ryan-Weber}, E.~V., {Garel}, T., \& {D{\'{\i}}az}, C.~G. 2014,
  \mnras, 441, 837

\bibitem[{{Cowie} {et~al.}(2009){Cowie}, {Barger}, \& {Trouille}}]{Cowie:2009}
{Cowie}, L.~L., {Barger}, A.~J., \& {Trouille}, L. 2009, \apj, 692, 1476

\bibitem[{{Crighton} {et~al.}(2015){Crighton}, {Murphy}, {Prochaska},
  {Worseck}, {Rafelski}, {Becker}, {Ellison}, {Fumagalli}, {Lopez}, {Meiksin},
  \& {O'Meara}}]{Crighton:2015}
{Crighton}, N.~H.~M., {Murphy}, M.~T., {Prochaska}, J.~X., {et~al.} 2015,
  \mnras, 452, 217

\bibitem[{{Danforth} {et~al.}(2016){Danforth}, {Keeney}, {Tilton}, {Shull},
  {Stocke}, {Stevans}, {Pieri}, {Savage}, {France}, {Syphers}, {Smith},
  {Green}, {Froning}, {Penton}, \& {Osterman}}]{Danforth:2016}
{Danforth}, C.~W., {Keeney}, B.~A., {Tilton}, E.~M., {et~al.} 2016, \apj, 817,
  111

\bibitem[{{Deharveng} {et~al.}(1997){Deharveng}, {Faiesse}, {Milliard}, \& {Le
  Brun}}]{Deharveng:1997}
{Deharveng}, J.-M., {Faiesse}, S., {Milliard}, B., \& {Le Brun}, V. 1997, \aap,
  325, 1259

\bibitem[{{Dove} \& {Shull}(1994)}]{Dove:1994}
{Dove}, J.~B., \& {Shull}, J.~M. 1994, \apj, 430, 222

\bibitem[{{Dove} {et~al.}(2000){Dove}, {Shull}, \& {Ferrara}}]{Dove:2000}
{Dove}, J.~B., {Shull}, J.~M., \& {Ferrara}, A. 2000, \apj, 531, 846

\bibitem[{{Fan} {et~al.}(2006){Fan}, {Carilli}, \& {Keating}}]{Fan:2006}
{Fan}, X., {Carilli}, C.~L., \& {Keating}, B. 2006, \araa, 44, 415

\bibitem[{{Fernandez-Soto} {et~al.}(2003){Fernandez-Soto}, {Lanzetta}, \&
  {Chen}}]{Fernandez-Soto:2003}
{Fernandez-Soto}, A., {Lanzetta}, K.~M., \& {Chen}, H.-W. 2003, \mnras, 342,
  1215

\bibitem[{{Finkelstein} {et~al.}(2015){Finkelstein}, {Ryan}, {Papovich},
  {Dickinson}, {Song}, {Somerville}, {Ferguson}, {Salmon}, {Giavalisco},
  {Koekemoer}, {Ashby}, {Behroozi}, {Castellano}, {Dunlop}, {Faber}, {Fazio},
  {Fontana}, {Grogin}, {Hathi}, {Jaacks}, {Kocevski}, {Livermore}, {McLure},
  {Merlin}, {Mobasher}, {Newman}, {Rafelski}, {Tilvi}, \&
  {Willner}}]{Finkelstein:2015}
{Finkelstein}, S.~L., {Ryan}, Jr., R.~E., {Papovich}, C., {et~al.} 2015, \apj,
  810, 71

\bibitem[{{Fumagalli} {et~al.}(2013){Fumagalli}, {O'Meara}, {Prochaska}, \&
  {Worseck}}]{Fumagalli:2013}
{Fumagalli}, M., {O'Meara}, J.~M., {Prochaska}, J.~X., \& {Worseck}, G. 2013,
  \apj, 775, 78

\bibitem[{{Gaikwad} {et~al.}(2017){Gaikwad}, {Khaire}, {Choudhury}, \&
  {Srianand}}]{Gaikwad:2017}
{Gaikwad}, P., {Khaire}, V., {Choudhury}, T.~R., \& {Srianand}, R. 2017,
  \mnras, 466, 838

\bibitem[{{Gnedin} {et~al.}(2008){Gnedin}, {Kravtsov}, \& {Chen}}]{Gnedin:2008}
{Gnedin}, N.~Y., {Kravtsov}, A.~V., \& {Chen}, H.-W. 2008, \apj, 672, 765

\bibitem[{{Hui} \& {Rutledge}(1999)}]{Hui:1999}
{Hui}, L., \& {Rutledge}, R.~E. 1999, \apj, 517, 541

\bibitem[{{Inoue} \& {Iwata}(2008)}]{Inoue:2008}
{Inoue}, A.~K., \& {Iwata}, I. 2008, \mnras, 387, 1681

\bibitem[{{Inoue} {et~al.}(2014){Inoue}, {Shimizu}, {Iwata}, \&
  {Tanaka}}]{Inoue:2014}
{Inoue}, A.~K., {Shimizu}, I., {Iwata}, I., \& {Tanaka}, M. 2014, \mnras, 442,
  1805

\bibitem[{{Izotov} {et~al.}(2016{\natexlab{a}}){Izotov}, {Orlitov{\'a}},
  {Schaerer}, {Thuan}, {Verhamme}, {Guseva}, \& {Worseck}}]{Izotov:2016a}
{Izotov}, Y.~I., {Orlitov{\'a}}, I., {Schaerer}, D., {et~al.}
  2016{\natexlab{a}}, \nat, 529, 178

\bibitem[{{Izotov} {et~al.}(2016{\natexlab{b}}){Izotov}, {Schaerer}, {Thuan},
  {Worseck}, {Guseva}, {Orlitov{\'a}}, \& {Verhamme}}]{Izotov:2016b}
{Izotov}, Y.~I., {Schaerer}, D., {Thuan}, T.~X., {et~al.} 2016{\natexlab{b}},
  \mnras, 461, 3683

\bibitem[{{Kim} {et~al.}(2013){Kim}, {Partl}, {Carswell}, \&
  {M{\"u}ller}}]{Kim:2013}
{Kim}, T.-S., {Partl}, A.~M., {Carswell}, R.~F., \& {M{\"u}ller}, V. 2013,
  \aap, 552, A77

\bibitem[{{Kollmeier} {et~al.}(2014){Kollmeier}, {Weinberg}, {Oppenheimer},
  {Haardt}, {Katz}, {Dav{\'e}}, {Fardal}, {Madau}, {Danforth}, {Ford},
  {Peeples}, \& {McEwen}}]{Kollmeier:2014}
{Kollmeier}, J.~A., {Weinberg}, D.~H., {Oppenheimer}, B.~D., {et~al.} 2014,
  \apjl, 789, L32

\bibitem[{{Leitet} {et~al.}(2013){Leitet}, {Bergvall}, {Hayes}, {Linn{\'e}}, \&
  {Zackrisson}}]{Leitet:2013}
{Leitet}, E., {Bergvall}, N., {Hayes}, M., {Linn{\'e}}, S., \& {Zackrisson}, E.
  2013, \aap, 553, A106

\bibitem[{{Leitherer} {et~al.}(2014){Leitherer}, {Ekstr{\"o}m}, {Meynet},
  {Schaerer}, {Agienko}, \& {Levesque}}]{Leitherer:2014}
{Leitherer}, C., {Ekstr{\"o}m}, S., {Meynet}, G., {et~al.} 2014, \apjs, 212, 14

\bibitem[{{Leitherer} {et~al.}(2016){Leitherer}, {Hernandez}, {Lee}, \&
  {Oey}}]{Leitherer:2016}
{Leitherer}, C., {Hernandez}, S., {Lee}, J.~C., \& {Oey}, M.~S. 2016, \apj,
  823, 64

\bibitem[{{Leitherer} {et~al.}(1999){Leitherer}, {Schaerer}, {Goldader},
  {Delgado}, {Robert}, {Kune}, {de Mello}, {Devost}, \&
  {Heckman}}]{Leitherer:1999}
{Leitherer}, C., {Schaerer}, D., {Goldader}, J.~D., {et~al.} 1999, \apjs, 123,
  3

\bibitem[{{Madau}(1995)}]{Madau:1995}
{Madau}, P. 1995, \apj, 441, 18

\bibitem[{{Madau} \& {Haardt}(2015)}]{Madau:2015}
{Madau}, P., \& {Haardt}, F. 2015, \apjl, 813, L8

\bibitem[{{Madau} {et~al.}(1999){Madau}, {Haardt}, \& {Rees}}]{Madau:1999}
{Madau}, P., {Haardt}, F., \& {Rees}, M.~J. 1999, \apj, 514, 648

\bibitem[{{McCandliss}(2003)}]{McCandliss:2003}
{McCandliss}, S.~R. 2003, \pasp, 115, 651

\bibitem[{{McCandliss} {et~al.}(2010){McCandliss}, {France}, {Osterman},
  {Green}, {McPhate}, \& {Wilkinson}}]{McCandliss:2010}
{McCandliss}, S.~R., {France}, K., {Osterman}, S., {et~al.} 2010, \apjl, 709,
  L183

\bibitem[{{Morrissey} {et~al.}(2005){Morrissey}, {Schiminovich}, {Barlow},
  {Martin}, {Blakkolb}, {Conrow}, {Cooke}, {Erickson}, {Fanson}, {Friedman},
  {Grange}, {Jelinsky}, {Lee}, {Liu}, {Mazer}, {McLean}, {Milliard}, {Randall},
  {Schmitigal}, {Sen}, {Siegmund}, {Surber}, {Vaughan}, {Viton}, {Welsh},
  {Bianchi}, {Byun}, {Donas}, {Forster}, {Heckman}, {Lee}, {Madore}, {Malina},
  {Neff}, {Rich}, {Small}, {Szalay}, \& {Wyder}}]{Morrissey:2005}
{Morrissey}, P., {Schiminovich}, D., {Barlow}, T.~A., {et~al.} 2005, \apjl,
  619, L7

\bibitem[{{Naidu} {et~al.}(2016){Naidu}, {Oesch}, {Reddy}, {Holden}, {Steidel},
  {Montes}, {Atek}, {Bouwens}, {Carollo}, {Cibinel}, {Illingworth}, {Labbe},
  {Magee}, {Morselli}, {Nelson}, {van Dokkum}, \& {Wilkins}}]{Naidu:2016}
{Naidu}, R.~P., {Oesch}, P.~A., {Reddy}, N., {et~al.} 2016, ArXiv e-prints,
  arXiv:1611.07038

\bibitem[{{Noterdaeme} {et~al.}(2012){Noterdaeme}, {Petitjean}, {Carithers},
  {P{\^a}ris}, {Font-Ribera}, {Bailey}, {Aubourg}, {Bizyaev}, {Ebelke},
  {Finley}, {Ge}, {Malanushenko}, {Malanushenko}, {Miralda-Escud{\'e}},
  {Myers}, {Oravetz}, {Pan}, {Pieri}, {Ross}, {Schneider}, {Simmons}, \&
  {York}}]{Noterdaeme:2012}
{Noterdaeme}, P., {Petitjean}, P., {Carithers}, W.~C., {et~al.} 2012, \aap,
  547, L1

\bibitem[Oke \& Gunn(1983)]{Oke:1983} Oke, J.~B., \& Gunn, J.~E.\ 1983, \apj, 266, 713 

\bibitem[{{O'Meara} {et~al.}(2013){O'Meara}, {Prochaska}, {Worseck}, {Chen}, \&
  {Madau}}]{O'Meara:2013}
{O'Meara}, J.~M., {Prochaska}, J.~X., {Worseck}, G., {Chen}, H.-W., \& {Madau},
  P. 2013, \apj, 765, 137

\bibitem[{{Paresce} {et~al.}(1980){Paresce}, {McKee}, \&
  {Bowyer}}]{Paresce:1980}
{Paresce}, F., {McKee}, C.~F., \& {Bowyer}, S. 1980, \apj, 240, 387

\bibitem[{{Prochaska} {et~al.}(2005){Prochaska}, {Herbert-Fort}, \&
  {Wolfe}}]{Prochaska:2005}
{Prochaska}, J.~X., {Herbert-Fort}, S., \& {Wolfe}, A.~M. 2005, \apj, 635, 123

\bibitem[{{Prochaska} {et~al.}(2014){Prochaska}, {Madau}, {O'Meara}, \&
  {Fumagalli}}]{Prochaska:2014}
{Prochaska}, J.~X., {Madau}, P., {O'Meara}, J.~M., \& {Fumagalli}, M. 2014,
  \mnras, 438, 476

\bibitem[{{Prochaska} {et~al.}(2010){Prochaska}, {O'Meara}, \&
  {Worseck}}]{Prochaska:2010}
{Prochaska}, J.~X., {O'Meara}, J.~M., \& {Worseck}, G. 2010, \apj, 718, 392

\bibitem[{{Razoumov} \& {Sommer-Larsen}(2010)}]{Razoumov:2010}
{Razoumov}, A.~O., \& {Sommer-Larsen}, J. 2010, \apj, 710, 1239

\bibitem[{{Reddy} {et~al.}(2016){Reddy}, {Steidel}, {Pettini},
  {Bogosavljevi{\'c}}, \& {Shapley}}]{Reddy:2016}
{Reddy}, N.~A., {Steidel}, C.~C., {Pettini}, M., {Bogosavljevi{\'c}}, M., \&
  {Shapley}, A.~E. 2016, \apj, 828, 108

\bibitem[{{Redwine} {et~al.}(2016){Redwine}, {McCandliss}, {Zheng}, {Fleming},
  {France}, {Osterman}, {Howk}, {Anderson}, \& {G{\"a}ensicke}}]{Redwine:2016}
{Redwine}, K., {McCandliss}, S.~R., {Zheng}, W., {et~al.} 2016, \pasp, 128,
  105006

\bibitem[{{Ribaudo} {et~al.}(2011){Ribaudo}, {Lehner}, \&
  {Howk}}]{Ribaudo:2011}
{Ribaudo}, J., {Lehner}, N., \& {Howk}, J.~C. 2011, \apj, 736, 42

\bibitem[{{Ricotti} {et~al.}(2002){Ricotti}, {Gnedin}, \&
  {Shull}}]{Ricotti:2002}
{Ricotti}, M., {Gnedin}, N.~Y., \& {Shull}, J.~M. 2002, \apj, 575, 49

\bibitem[{{Robertson} {et~al.}(2015){Robertson}, {Ellis}, {Furlanetto}, \&
  {Dunlop}}]{Robertson:2015}
{Robertson}, B.~E., {Ellis}, R.~S., {Furlanetto}, S.~R., \& {Dunlop}, J.~S.
  2015, \apjl, 802, L19

\bibitem[{{Schechter}(1976)}]{Schechter:1976}
{Schechter}, P. 1976, \apj, 203, 297

\bibitem[{{Scott} {et~al.}(2004){Scott}, {Kriss}, {Brotherton}, {Green},
  {Hutchings}, {Shull}, \& {Zheng}}]{Scott:2004}
{Scott}, J.~E., {Kriss}, G.~A., {Brotherton}, M., {et~al.} 2004, \apj, 615, 135

\bibitem[{{Shapley} {et~al.}(2006){Shapley}, {Steidel}, {Pettini},
  {Adelberger}, \& {Erb}}]{Shapley:2006}
{Shapley}, A.~E., {Steidel}, C.~C., {Pettini}, M., {Adelberger}, K.~L., \&
  {Erb}, D.~K. 2006, \apj, 651, 688

\bibitem[{{Shapley} {et~al.}(2016){Shapley}, {Steidel}, {Strom},
  {Bogosavljevi{\'c}}, {Reddy}, {Siana}, {Mostardi}, \& {Rudie}}]{Shapley:2016}
{Shapley}, A.~E., {Steidel}, C.~C., {Strom}, A.~L., {et~al.} 2016, \apjl, 826,
  L24

\bibitem[{{Sharma} {et~al.}(2016){Sharma}, {Theuns}, {Frenk}, {Bower}, {Crain},
  {Schaller}, \& {Schaye}}]{Sharma:2016}
{Sharma}, M., {Theuns}, T., {Frenk}, C., {et~al.} 2016, \mnras, 458, L94

\bibitem[{{Shull} {et~al.}(2015){Shull}, {Moloney}, {Danforth}, \&
  {Tilton}}]{Shull:2015}
{Shull}, J.~M., {Moloney}, J., {Danforth}, C.~W., \& {Tilton}, E.~M. 2015,
  \apj, 811, 3

\bibitem[{{Siana} {et~al.}(2007){Siana}, {Teplitz}, {Colbert}, {Ferguson},
  {Dickinson}, {Brown}, {Conselice}, {de Mello}, {Gardner}, {Giavalisco}, \&
  {Menanteau}}]{Siana:2007}
{Siana}, B., {Teplitz}, H.~I., {Colbert}, J., {et~al.} 2007, \apj, 668, 62

\bibitem[{{Smith} {et~al.}(1981){Smith}, {Carswell}, {Whelan}, {Wilkes},
  {Boksenberg}, {Clowes}, {Savage}, {Cannon}, \& {Wall}}]{Smith:1981}
{Smith}, M.~G., {Carswell}, R.~F., {Whelan}, J.~A.~J., {et~al.} 1981, \mnras,
  195, 437

\bibitem[{{Smith} {et~al.}(1996){Smith}, {Esmond}, {Heise}, \&
  {Kurucz}}]{Smith:1996}
{Smith}, P.~L., {Esmond}, J.~R., {Heise}, C., \& {Kurucz}, R.~L. 1996, in UV
  and X-ray Spectroscopy of Astrophysical and Laboratory Plasmas, ed.
  K.~{Yamashita} \& T.~{Watanabe}, 513--516

\bibitem[{{Steidel} {et~al.}(2001){Steidel}, {Pettini}, \&
  {Adelberger}}]{Steidel:2001}
{Steidel}, C.~C., {Pettini}, M., \& {Adelberger}, K.~L. 2001, \apj, 546, 665

\bibitem[{{Steidel} {et~al.}(1995){Steidel}, {Pettini}, \&
  {Hamilton}}]{Steidel:1995}
{Steidel}, C.~C., {Pettini}, M., \& {Hamilton}, D. 1995, \aj, 110, 2519

\bibitem[{{Verner} {et~al.}(1996){Verner}, {Ferland}, {Korista}, \&
  {Yakovlev}}]{Verner:1996}
{Verner}, D.~A., {Ferland}, G.~J., {Korista}, K.~T., \& {Yakovlev}, D.~G. 1996,
  \apj, 465, 487

\bibitem[{{Weingartner} \& {Draine}(2001)}]{Weingartner:2001}
{Weingartner}, J.~C., \& {Draine}, B.~T. 2001, \apj, 548, 296

\bibitem[{{Weymann} {et~al.}(1998){Weymann}, {Jannuzi}, {Lu}, {Bahcall},
  {Bergeron}, {Boksenberg}, {Hartig}, {Kirhakos}, {Sargent}, {Savage},
  {Schneider}, {Turnshek}, \& {Wolfe}}]{Weymann:1998}
{Weymann}, R.~J., {Jannuzi}, B.~T., {Lu}, L., {et~al.} 1998, \apj, 506, 1

\bibitem[{{Wise} {et~al.}(2014){Wise}, {Demchenko}, {Halicek}, {Norman},
  {Turk}, {Abel}, \& {Smith}}]{Wise:2014}
{Wise}, J.~H., {Demchenko}, V.~G., {Halicek}, M.~T., {et~al.} 2014, \mnras,
  442, 2560

\bibitem[{{Worseck} {et~al.}(2014){Worseck}, {Prochaska}, {O'Meara}, {Becker},
  {Ellison}, {Lopez}, {Meiksin}, {M{\'e}nard}, {Murphy}, \&
  {Fumagalli}}]{Worseck:2014}
{Worseck}, G., {Prochaska}, J.~X., {O'Meara}, J.~M., {et~al.} 2014, \mnras,
  445, 1745

\bibitem[{{Xu} {et~al.}(2016){Xu}, {Wise}, {Norman}, {Ahn}, \&
  {O'Shea}}]{Xu:2016}
{Xu}, H., {Wise}, J.~H., {Norman}, M.~L., {Ahn}, K., \& {O'Shea}, B.~W. 2016,
  \apj, 833, 84
  
\bibitem[{{Yajima} {et~al.}(2011){Yajima}, {Choi}, \& {Nagamine}}]{Yajima:2011}
{Yajima}, H., {Choi}, J.-H., \& {Nagamine}, K. 2011, \mnras, 412, 411

\bibitem[{{Yajima} {et~al.}(2014){Yajima}, {Li}, {Zhu}, {Abel}, {Gronwall}, \&
  {Ciardullo}}]{Yajima:2014}
{Yajima}, H., {Li}, Y., {Zhu}, Q., {et~al.} 2014, \mnras, 440, 776

\bibitem[{{Yoshida} {et~al.}(2006){Yoshida}, {Shimasaku}, {Kashikawa}, {Ouchi},
  {Okamura}, {Ajiki}, {Akiyama}, {Ando}, {Aoki}, {Doi}, {Furusawa},
  {Hayashino}, {Iwamuro}, {Iye}, {Karoji}, {Kobayashi}, {Kodaira}, {Kodama},
  {Komiyama}, {Malkan}, {Matsuda}, {Miyazaki}, {Mizumoto}, {Morokuma},
  {Motohara}, {Murayama}, {Nagao}, {Nariai}, {Ohta}, {Sasaki}, {Sato},
  {Sekiguchi}, {Shioya}, {Tamura}, {Taniguchi}, {Umemura}, {Yamada}, \&
  {Yasuda}}]{Yoshida:2006}
{Yoshida}, M., {Shimasaku}, K., {Kashikawa}, N., {et~al.} 2006, \apj, 653, 988

\end{thebibliography}
\end{document}